\documentclass[aps,pra,showpacs,twocolumn,amsmath,amssymb,superscriptaddress,nofootinbib]{revtex4}
\usepackage{graphicx}
\usepackage{bm}
\usepackage{amssymb}
\usepackage{amsmath}
\usepackage{latexsym}
\usepackage{color}
\usepackage{soul}
\usepackage[normalem]{ulem}
\newcommand{\vc}[1]{\mbox{\boldmath $#1$}} 
\newcommand{\ind}[1]{_{#1}}    
\newcommand{\indrm}[1]{_{\mathrm {#1}}}    
\newcommand{\dirate}{{\mathcal D}}   
\newcommand{\sgn}{s}   

\newcommand{\fspace}{P}   
\newcommand{\thinlens}{L}   
\newcommand{\crystal}{C}   
\newcommand{\crystalsp}{K}

\newcommand{\focus}{F}   
\newcommand{\dcomm}[1]{b_{\cup_{\ind{#1}}}}   
\newcommand{\acomm}[1]{1/b_{\cup_{\ind{#1}}}}
\newcommand{\bcomm}[1]{B_{\cup_{\ind{#1}}}}      
\newcommand{\gcomm}[1]{G_{\cup_{\ind{#1}}}}      
\newcommand{\fcomm}[1]{\dirate_{\cup_{\ind{#1}}}}
\newcommand{\elmt}{e}   
\newcommand{\cmb}{C}   
\newcommand{\an}{R}   
\newcommand{\mo}{D}   
\newcommand{\etra}{\varepsilon}
\newcommand{\deis}[1]{\Delta E_{\indrm{#1}}^{({\mathrm s})}}   
\newcommand{\dais}[1]{\Delta \theta_{\indrm{#1}}^{({\mathrm s})}}   
\newcommand{\ysize}{\Delta Y}   
\newcommand{\xsize}{\Delta x_{\ind{1}}}   
\newcommand{\Xsize}{\Delta X}   
\newcommand{\beamsize}{B}   
\newcommand{\xphisize}{\Delta x_{\ind{1\varphi}}}   
\newcommand{\esi}{XES1}    
\newcommand{\esoi}{XES01}    
\newcommand{\dband}{\Delta E_{\indrm{D}}} 
\newcommand{\rband}{\Delta E_{\indrm{R}}} 
\newcommand{\xband}{\Delta E_{\indrm{X}}} 
\newcommand{\band}{\Delta E} 
\newcommand{\raccept}{\Delta\theta_{\ind{\an}}} 
\newcommand{\daccept}{\Delta\theta_{\ind{\mo}}} 
\newcommand{\xaccept}{\Delta\theta_{\ind{X}}} 
\newcommand{\rdtheta}{\Delta\theta_{\ind{\an}}^{\prime}} 
\newcommand{\ddtheta}{\Delta\theta_{\ind{\mo}}^{\prime}} 
\newcommand{\xdtheta}{\Delta\theta_{\ind{X}}^{\prime}} 
\begin{document}  
\title{Theory and optical design of x-ray echo spectrometers}
\author{Yuri Shvyd'ko}
\email{shvydko@aps.anl.gov}
\affiliation{Advanced Photon Source, Argonne National Laboratory,   Argonne, Illinois 60439, USA} 
\begin{abstract} 
X-ray echo spectroscopy, a space-domain counterpart of neutron spin
echo, is a recently proposed inelastic x-ray scattering (IXS)
technique.  X-ray echo spectroscopy relies on imaging IXS spectra, and
does not require x-ray monochromatization. Due to this, the echo-type
IXS spectrometers are broadband, and thus have a potential to
simultaneously provide dramatically increased signal strength, reduced
measurement times, and higher resolution compared to the traditional
narrow-band scanning-type IXS spectrometers. The theory of x-ray echo
spectrometers presented in \cite{Shvydko16} is developed here further
with a focus on questions of practical importance, which could
facilitate optical design and assessment of the feasibility and
performance of the echo spectrometers.  Among others, the following
questions are addressed: spectral resolution, refocusing condition,
echo spectrometer tolerances, refocusing condition adjustment,
effective beam size on the sample, spectral window of imaging and
scanning range, impact of the secondary source size on the spectral
resolution, angular dispersive optics, focusing and collimating
optics, and detector's spatial resolution. Examples of optical
designs and characteristics of echo spectrometers with 1-meV and
0.1-meV resolutions are presented.
\end{abstract}
%
\maketitle

\section{Introduction}

Photon and neutron inelastic scattering spectrometers are microscopes
for imaging condensed matter dynamics at very small length and time
scales.  Momentum-resolved inelastic x-ray scattering (IXS) is a
technique introduced \cite{BDP87,Burkel} and widely used
\cite{Sette98,Burkel2,KS07,MonacoIXS15,Baron16} at storage-ring-based
synchrotron radiation facilities. Despite numerous advances, progress
on many of the key problems in condensed matter physics is held back
because current inelastic scattering probes are limited in energy
$\Delta \varepsilon$, momentum $\Delta Q$ resolution, and in signal
strength.  The signal strength is limited by several factors. First,
undulator spectral flux is at the limit of what is possible with
current storage-ring-based x-ray source
technology. High-repetition-rate self-seeded x-ray free-electron
lasers in the future may provide orders of magnitude more spectral
flux than what is possible at storage ring sources, and therefore may
substantially improve IXS signal strength \cite{CGk16}.  Second,
because the signal strength $S \propto \Delta \varepsilon^2 \Delta
Q^2$ scales quadratically with the spectral and momentum transfer
resolutions of traditional IXS instruments, it is severely limited by
the small values of $\Delta \varepsilon$ and $\Delta Q$ required for
IXS.  For example, improving the resolution by an order of magnitude
from the currently available $\Delta \varepsilon=1.5$~meV and $\Delta
Q=1.5$~nm$^{-1}$ to a very much desired $\Delta \varepsilon=0.1$~meV
and $\Delta Q=0.1$~nm$^{-1}$ should inevitably result in a four orders
of magnitude signal reduction. Such improvements in the resolutions of
traditional IXS instruments seem, therefore, to be impractical at
least at storage-ring-based x-ray sources.

A recently proposed x-ray echo spectroscopy technique can change the
situation dramatically and open up completely new opportunities
\cite{Shvydko16}.  The essential features of echo spectroscopy are,
first, that it relies on imaging IXS spectra and, second, that it does
not require x-ray monochromatization, as conventional IXS
spectrometers do. Due to this, the echo-type IXS spectrometers may be
broadband devices, and therefore have a potential to simultaneously
provide dramatically increased signal strength, reduced measurement
times, and practical measurements having higher resolution.

In the present paper, we develop further the theory of the x-ray echo
spectrometers with a focus on questions of practical importance, which
could help in optical design and in assessing the feasibility and
performance of echo spectrometers.  Among others, the following
questions are addressed: spectral resolution, refocusing condition,
echo spectrometer tolerances, refocusing condition adjustment,
effective beam size on the sample, spectral window of imaging and
scanning range, impact of secondary source size on the spectral
resolution, angular dispersive optics, focusing and
collimating optics, and detector's spatial resolution.

Examples of optical designs and characteristics of x-ray echo
spectrometers with 1-meV and 0.1-meV resolutions are presented and
supported by the theory. In particular, the echo-type
0.1-meV-resolution IXS spectrometer is predicted to feature the same
signal strength, however, a 10 times improved spectral resolution and
a 25 times improved momentum transfer resolution (0.05~nm$^{-1}$)
compared to a state-of-the-art narrow-band scanning-type 1-meV and
1-nm$^{-1}$ resolution IXS spectrometer \cite{TAS11,SSD11}.

\section{Basic theory and principal scheme}

\begin{figure}[t!]
\includegraphics[width=0.5\textwidth]{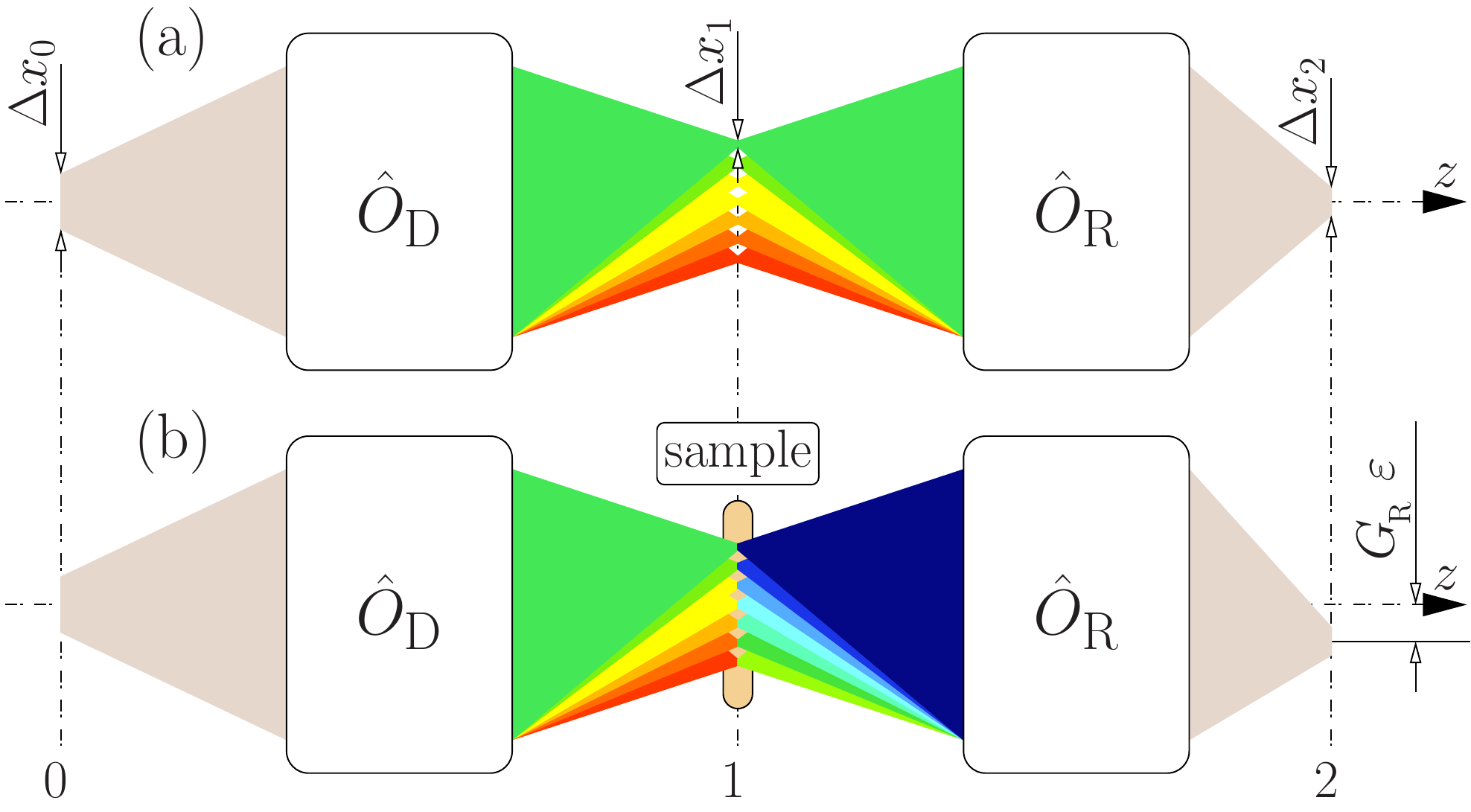}
\caption{Graphical presentation of the echo spectroscopy
  principles. (a) Photons from a source with a linear size $\Delta
  x_{\indrm{0}}$ in reference source plane $0$ are focused onto a
  spot $\Delta x_{\indrm{1}}$ in the intermediate image plane $1$ by a
  focusing-dispersing system $\hat{O}_{\indrm{\mo}}$. Each spectral
  component, indicated by different color, is
  focused at a different location due to dispersion in
  $\hat{O}_{\indrm{\mo}}$. All spectral components of the x~rays are
  refocused by a consecutive time-reversal focusing-dispersing system
  $\hat{O}_{\indrm{\an}}$ onto the same spot $\Delta x_{\indrm{2}}$
  (echo) in the image plane $2$. (b) Inelastic x-ray scattering with
  an energy transfer $\etra$ (indicated by changed color) 
  from a sample in reference plane $1$ results in
  a lateral shift $G_{\indrm{\an}} \etra$ of the echo signal
  equal for all spectral components.
}
\label{fig001}
\end{figure}

We start by considering optical systems featuring a combination of
focusing and energy dispersing capabilities. We assume that such
systems can, first, focus {\em monochromatic} x~rays from a source of
a linear size $\Delta x_{\ind{0}}$ in a source plane (reference plane
$0$ perpendicular to the optical axis $z$ in Fig.~\ref{fig001}) onto
an intermediate image plane (reference plane $1$ in Fig.~\ref{fig001})
with an image linear size $\Delta x_{\ind{1}}= |A| \Delta
x_{\ind{0}}$, where $A$ is a magnification factor of the optical
system. In addition, the system can disperse photons in such a way
that the location of the image for photons with an energy $E+\delta E$
is displaced in the image plane by $G \delta E$ from the location of
the image for photons with energy $E$. Here, $G$ is a linear
dispersion rate of the system, which is a product of the angular
dispersion rate, hereafter denoted as $\dirate$, and a characteristic
distance to the image plane.  As a result, although monochromatic
x~rays are focused, the whole spectrum of x~rays is defocused, due to
linear dispersion.

We will use the ray-transfer matrix technique
\cite{KL66,MK80-1,Siegman} to propagate paraxial x~rays through such
optical systems and to determine linear and angular sizes of the x-ray
beams along the optical axis. A paraxial ray in any reference plane is
characterized by its distance $x$ from the optical axis, by its angle
$\xi$ with respect to that axis, and the deviation $\delta E$ of the
photon energy from a nominal value $E$.  The ray vector
$\vc{r}_{\ind{0}}=(x_{\ind{0}},\xi_{\ind{0}},\delta E)$ at an input
source plane is transformed to
$\vc{r}_{\ind{1}}=(x_{\ind{1}},\xi_{\ind{1}},\delta
E)=\hat{O}\vc{r}_{\ind{0}}$ at the output reference plane (image
plane), where $\hat{O}=\{ABG;CDF;001\}$ is a ray-transfer matrix of an
optical element placed between the planes.  Only elastic processes in
the optical systems are taken into account; this is reflected by zero
and unity terms in the lowest row of the ray-transfer matrices.

Focusing of the monochromatic spectral components requires that matrix
element $B=0$. The ray-transfer matrix of any focusing-dispersing
system in a general case therefore reads as
\begin{equation}
\hat{O}  = 
 \{A \,0\,G;\,CDF;001\}
\label{matrix}
\end{equation}
with $A$ and $G$ elements defined above. The system blurs the
polychromatic source image, because of linear dispersion, as mentioned
earlier and graphically presented in Fig.~\ref{fig001}(a).  However,
another focusing-dispersing system
can be used to
refocus the source onto reference plane 2.  Indeed, propagation of
x~rays through the defocusing system $\hat{O}_{\indrm{\mo}}$ and a
second system, which we will refer to as a refocusing or time-reversal
system $\hat{O}_{\indrm{\an}}$ (see Fig.~\ref{fig001}), is given by a
combined ray-transfer matrix 
\begin{multline}
\hat{O}_{\indrm{\cmb}}  = \hat{O}_{\indrm{\an}} \hat{O}_{\indrm{\mo}} = \{A_{\indrm{\cmb}}\,0\,G_{\indrm{\cmb}};C_{\indrm{\cmb}}D_{\indrm{\cmb}}F_{\indrm{\cmb}};001\}\\  = \left(\!\! \begin{array}{ccc} A_{\indrm{\an}} A_{\indrm{\mo}}  & 0 &  A_{\indrm{\an}}G_{\indrm{\mo}}+G_{\indrm{\an}}  \\ C_{\indrm{\an}}A_{\indrm{\mo}}\!+\!D_{\indrm{\an}}C_{\indrm{\mo}} 
  & D_{\indrm{\an}} D_{\indrm{\mo}}  & C_{\indrm{\an}}G_{\indrm{\mo}}\!+\!D_{\indrm{\an}}F_{\indrm{\mo}}\!+\!F_{\indrm{\an}} \\ 0 & 0 & 1 \end{array}\!\! \right),
\label{comb}
\end{multline}
and by a ray vector $\vc{r}_{\ind{2}}=(x_{\ind{2}},\xi_{\ind{2}},\delta
E)=\hat{O}_{\indrm{\cmb}}\vc{r}_{\ind{0}}$. 

Here we arrive at a crucial point. If 
\begin{equation}
G_{\indrm{\cmb}}=A_{\indrm{\an}}G_{\indrm{\mo}}+G_{\indrm{\an}}=0,
\label{refocus}
\end{equation}
the linear dispersion at the exit of the combined system vanishes,
because dispersion in the defocusing system is compensated (time
reversed) by dispersion in the refocusing system. As a result, the
combined system refocuses all photons independent of the photon energy
to the same location, $x_{\ind{2}}$ in image plane $2$, to a spot with a
linear size
\begin{equation}
\Delta x_{\ind{2}}= |A_{\indrm{\an}} A_{\indrm{\mo}}| \Delta x_{\ind{0}}\equiv  |A_{\indrm{\an}} | \Delta x_{\ind{1}}, 
\label{magnification}
\end{equation}
as shown schematically in Fig.~\ref{fig001}(a).  Such behavior is an
analog of the echo phenomena \cite{Hahn50,Mezei80}. Here, however, it
takes place in space, rather than in the time domain.\footnote{It is
  noteworthy that angular dispersion always results in an inclined
  intensity front, i.e., in dispersion both perpendicular to and along
  the beam propagation direction \cite{SL12}.  Therefore, x~rays are
  defocused and refocused also in the time domain, as in spin-echo. As
  a result, inelastic scattering spectra can be also mapped by
  measuring time distributions in the detector, given a short-pulse
  source.}

Now, what happens if a sample is placed into the intermediate image
plane $1$ [Fig.~\ref{fig001}(b)], which can scatter photons
inelastically?  In an inelastic scattering process, a photon with an
arbitrary energy $E+\delta E$ changes its value to $E+\delta E+
\etra$. Here $\etra$ is an energy transfer in the inelastic scattering
process.  The ray vector
$\vc{r}_{\ind{1}}=(x_{\ind{1}},\xi_{\ind{1}},\delta E)$ before
scattering transforms to
$\vc{r}_{\ind{1}}^{\prime}=(x_{\ind{1}},\xi_{\ind{1}}^{\prime},\delta
E+\etra)$ after inelastic scattering. Propagation of
$\vc{r}_{\ind{1}}^{\prime}$ through the time-reversal system results
in a ray vector
$\vc{r}_{\ind{2}}^{\prime}=(x_{\ind{2}}^{\prime},\xi_{\ind{2}}^{\prime},\delta
E+\etra)= \hat{O}_{\indrm{\an}}\vc{r}_{\ind{1}}^{\prime}$.  Assuming
that refocusing condition \eqref{refocus} holds, we come to a decisive
point: all photons independent of the incident photon energy $E+\delta
E$ are refocused to the same location
\begin{equation}
x_{\ind{2}}^{\prime}=x_{\ind{2}}+G_{\indrm{\an}}  \etra,\hspace{0.5cm}  x_{\ind{2}}=A_{\indrm{\an}} A_{\indrm{\mo}} x_{\ind{o}},
\label{magnification2}
\end{equation}
which is, however, shifted from $x_{\ind{2}}$ by $G_{\indrm{\an}}
\etra$, a value proportional to the energy transfer $\etra$ in the
inelastic scattering process. The essential point is that the
combined defocusing-refocusing system maps the inelastic scattering
spectrum onto image plane $2$. The image is independent of the
spectral composition $E+\delta E$ of the photons in the incident beam.

The spectral resolution $\Delta \etra$ of the echo spectrometer is
calculated from the condition that the shift due to inelastic
scattering $x_{\ind{2}}^{\prime}-x_{\ind{2}}=G_{\indrm{\an}} 
\etra$ is at least as large as the linear size $\Delta
x_{\ind{2}}$ of the echo signal  in Eq.~\eqref{magnification}:
\begin{equation}
\Delta \etra =    \frac{\Delta x_{\ind{2}}}{|G_{\indrm{\an}}|}  \equiv  \frac{|A_{\indrm{\an}}|\Delta x_{\ind{1}} }{|G_{\indrm{\an}}|} \equiv  \frac{|A_{\indrm{\an}} A_{\indrm{\mo}}| \Delta x_{\ind{o}}}{|G_{\indrm{\an}}|}.
\label{resolution}
\end{equation}
Here it is assumed that the spatial resolution of an x-ray detector in
reference plane $2$ is better than $\Delta x_{\ind{2}}$.  

These results constitute the underlying principle of x-ray echo
spectroscopy. Most important is that the
x-ray echo spectroscopy technique involves imaging the inelastic
scattering spectrum without requiring x-ray monochromatization.

Perfect refocusing takes place if the linear dispersion of the
combined system
$G_{\indrm{\cmb}}=A_{\indrm{\an}}G_{\indrm{\mo}}+G_{\indrm{\an}}$
vanishes, as in Eq.~\eqref{refocus}. Refocusing can still take place
with good accuracy if $|G_{\indrm{\cmb}}|$ is sufficiently small
\begin{equation}
|G_{\indrm{\cmb}}| \band \ll \Delta x_{\ind{2}},
\label{tolerances} 
\end{equation}
and, therefore,  does not deteriorate the spectral resolution.
Here $\band$ is a spectral bandwidth of x~rays in each {\em particular}
point in image plane $2$. In the following, $\band$ will be referred
to as an effective bandwidth of the spectrometer. It should not be
confused with the spectral bandwidth $\dband$ of the defocusing system
or the spectral window of imaging $\rband$ of the refocusing
system. As discussed in Sec.~\ref{vspotsize}, $\band$ is typically
smaller than $\dband$ or $\rband$.  Tolerances on the echo
spectrometer parameters, on the sample shape, etc., can be calculated with
Eq.~\eqref{tolerances}, as discussed in more detail in Sec.~\ref{sec-tolerances}. 

The above approach is general and applicable to any frequency
domain. A particular version was proposed and realized in the soft
x-ray domain, with diffraction gratings as dispersing elements
\cite{Fung04,Lai14}.  Our focus is IXS in the hard x-ray
domain.\footnote{Dispersion compensation was also applied to IXS
  spectrometers in the hard x-ray regime
  \cite{SNM68,SCHULKE86,Schulke2007,SBD01}. Because Bragg's law
  dispersion was compensated, the spectral resolution of the
  spectrometers was limited (to $\simeq 1$~eV) by the Darwin widths of
  the Bragg reflections involved. The approach presented in this paper
  uses the angular dispersion, with the spectral resolution not
  limited by the Darwin width, and relies on broadband IXS spectra
  imaging.}  Diffraction gratings are not practical in the hard x-ray
regime.  However, the angular dispersion effect of the diffraction
grating can be achieved in the hard x-ray regime by Bragg diffraction
from {\em asymmetrically cut} crystals, i.e., from crystals with the
reflecting atomic planes not parallel to the entrance surface, as
demonstrated in \cite{Shvydko-SB,SLK06}. The crystals in asymmetric
Bragg diffraction are {\em the} hard x-ray analog of optical
diffraction gratings or optical prisms.  A large dispersion rate is a
key for achieving high spectral resolution in angular-dispersive x-ray
spectrometers \cite{SSS14,Shvydko15}, including echo spectrometers;
see Eq.~\eqref{resolution}.  In the following two steps, we will show
how the principal scheme of a generic echo spectrometer presented
above, can be realized in the hard x-ray regime.

\begin{figure}[t!]
\includegraphics[width=0.5\textwidth]{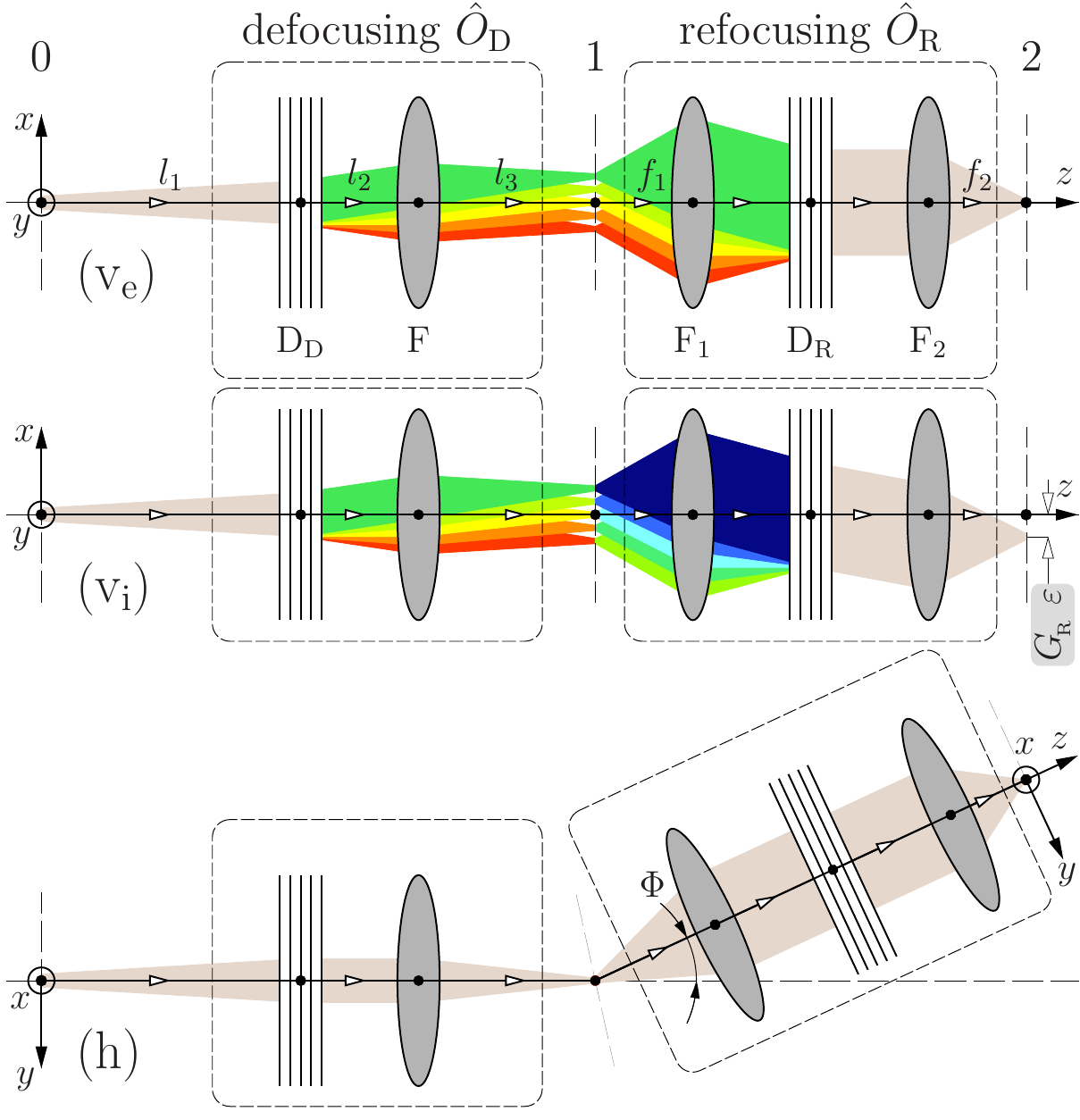}
\caption{Optical scheme of an x-ray echo spectrometer, composed of
  the defocusing $\hat{O}_{\indrm{\mo}}$ and refocusing
  $\hat{O}_{\indrm{\an}}$ dispersing systems, the x-ray source in
  reference plane $0$, the sample in $1$, and the position-sensitive
  detector in $2$.  The defocusing system $\hat{O}_{\indrm{\mo}}$
  consists of a dispersing Bragg diffracting (multi)crystal element
  D$_{\indrm{\mo}}$s and of a focusing element F.  The refocusing
  system $\hat{O}_{\indrm{\an}}$ is of a spectrograph type comprising
  collimating element F$_{\ind{1}}$, a dispersing Bragg diffracting
  (multi)crystal element D$_{\indrm{\an}}$, and an imaging element
  F$_{\ind{2}}$.  The spectrometer is shown in the vertical dispersion
  plane ($x,z$) for elastic (v$_{\indrm{e}}$) and inelastic
  (v$_{\indrm{i}}$) scattering, and in the horizontal scattering plane
  ($y,z$) with the refocusing system at a scattering angle $\Phi$ (h).
  $\Phi$ defines the momentum transfer $Q=2K\sin\Phi/2$ of a photon with
  momentum $K$.    }
\label{fig002}
\end{figure}

\section{Optical scheme}
\label{optical-scheme}

In the first step, we consider a more explicit optical scheme of the
hard x-ray echo spectrometer, shown in Fig.~\ref{fig002}, with the
defocusing $\hat{O}_{\indrm{\mo}}$ and refocusing
$\hat{O}_{\indrm{\an}}$ dispersing systems equipped
with specific optical elements. The x-ray source is in reference plane
$0$, the sample (secondary source) is in plane $1$, and the
position-sensitive detector is in plane $2$.  The defocusing system
$\hat{O}_{\indrm{\mo}}$ comprises a Bragg diffracting (multi)crystal
dispersing element D$_{\indrm{\mo}}$ and a focusing element F.  As has
been shown in \cite{Shvydko15}, see also Table~\ref{tab2} in Appendix~\ref{ray-transfer-matrices}, such a
system can be represented by a ray-transfer matrix \eqref{matrix} with
the magnification $A_{\indrm{\mo}}$ and linear dispersion
$G_{\indrm{\mo}}$ matrix elements given by
\begin{equation}
A_{\indrm{\mo}}\!=\! - \frac{1}{\dcomm{\mo}}\!\frac{l_{\ind{3}}}{l_{\ind{12}}},\hspace{0.20cm}
G_{\indrm{\mo}}\!=\!\fcomm{\mo}  \frac{l_{\ind{3}} l_{\ind{1}}}{\dcomm{\mo}^2 l_{\ind{12}}},\hspace{0.20cm} l_{\ind{12}}\! =\!\frac{l_{\ind{1}}}{\dcomm{\mo}^2} + l_{\ind{2}}. 
\label{defocusing}
\end{equation}
Here, $l_{\ind{1}}$, $l_{\ind{2}}$, and $l_{\ind{3}}$ are the
distances between the x-ray source, the dispersing element
D$_{\indrm{\mo}}$, the focusing element F with focal length
$f=(l_{\ind{12}}^{-1}+l_{\ind{3}}^{-1})^{-1}$, and the sample in the
image plane $1$, respectively (Fig.~\ref{fig002}). The dispersing
(multi)crystal system D$_{\indrm{\mo}}$ is characterized by the
cumulative angular dispersion rate $\fcomm{\mo}$ and cumulative
asymmetry factor $\dcomm{\mo}$, which are defined in \cite{Shvydko15}
(see also Sec.~\ref{angulardispersionrate} and Table~\ref{tab2} in
Appendix~\ref{ray-transfer-matrices}).

For the spectrometer to feature a large throughput, the refocusing
system $\hat{O}_{\indrm{\an}}$ has to be capable of collecting x-ray
photons scattered from the sample in a large solid angle. An example
of a focusing-dispersing system with a large solid acceptance angle is
schematically shown in Fig~\ref{fig002}. It is equivalent to the
spectrograph scheme discussed in \cite{Shvydko15}.  Collimating
element F$_{\ind{1}}$ with a focal distance $f_{\ind{1}}$ collects
photons in a large solid angle and makes x-ray beams of each spectral
component parallel. The collimated beam impinges upon the Bragg
(multi)crystal dispersing element D$_{\indrm{\an}}$ with the
cumulative angular dispersion rate $\fcomm{\an}$ and the cumulative
asymmetry factor $\dcomm{\an}$. Imaging element F$_{\ind{2}}$ with a
focal distance $f_{\ind{2}}$ focuses x~rays onto the
position-sensitive detector in image plane $2$.  As shown in
\cite{Shvydko15} (see also Table~\ref{tab2} in
Appendix~\ref{ray-transfer-matrices}), such a system is described by a
ray-transfer matrix \eqref{matrix} with the magnification
$A_{\indrm{\an}}$ and linear dispersion $G_{\indrm{\an}}$ matrix
elements given by
\begin{equation}
A_{\indrm{\an}}=-\frac{\dcomm{\an}f_{\ind{2}}\!}{f_{\ind{1}}\!},\hspace{0.5cm} G_{\indrm{\an}}=\fcomm{\an} f_{\ind{2}} .
\label{refocusing}
\end{equation}
Using Eqs.~\eqref{refocus}, \eqref{defocusing}, and \eqref{refocusing},
we obtain for the refocusing condition of the hard x-ray echo
spectrometer schematically presented in Fig.~\ref{fig002}
\begin{equation}
 \frac{l_{\ind{3}} l_{\ind{1}}}{l_{\ind{1}} + \dcomm{\mo}^2 l_{\ind{2}}} \fcomm{\mo} = f_{\ind{1}} \frac{\fcomm{\an}}{\dcomm{\an}}. 
\label{refocus2}
\end{equation}
The dispersing element D$_{\indrm{\mo}}$ can be placed from the source
at a large distance $l_{\ind{1}} \gg \dcomm{\mo}^2 l_{\ind{2}}$. In
this case, the refocusing condition \eqref{refocus2} reads
\begin{equation}
 l_{\ind{3}}\fcomm{\mo} \simeq  f_{\ind{1}} \frac{\fcomm{\an}}{\dcomm{\an}}. 
\label{refocus23}
\end{equation}
We note that for the refocusing condition to be fulfilled, 
$\fcomm{\mo}$ and  $\fcomm{\an}/\dcomm{\an}$ should have the same sign. 

For the spectral resolution $\Delta \etra$ of the hard x-ray echo
spectrometer schematically presented in Fig.~\ref{fig002}, we obtain
from Eqs.~\eqref{resolution}, \eqref{defocusing}, and
\eqref{refocusing}
\begin{equation}
\Delta \etra =   \frac{|\dcomm{\an}|}{|\fcomm{\an}|} \frac{\Delta x_{\ind{1}}}{f_{\ind{1}}}. 
\label{resolution2}
\end{equation}
As follows from Eq.~\eqref{resolution2}, the spectral resolution of
the echo spectrometer is defined solely by the parameters of the
refocusing system, and it is equivalent to the resolution of the hard
x-ray spectrograph \cite{Shvydko15}. As pointed out before, the
resolution is independent of the spectral composition of the x~rays
impinging on the sample.  The parameters of the defocusing system
determine only the size of the secondary monochromatic source on the
sample $\Delta x_{\ind{1}}=|A_{\indrm{\mo}}|\Delta x_{\ind{0}}$, with
$A_{\indrm{\mo}}$ defined in Eq.~\eqref{defocusing}.


Equation \eqref{resolution2} can be used to estimate the magnitude of
the dispersion rate of the dispersing element D$_{\indrm{\an}}$ or
more precisely  the ratio $|\fcomm{\an}/\dcomm{\an}|$ required to
achieve the desired spectral resolution. For example, for an x-ray
echo spectrometer with a resolution $\Delta \etra = 1$~meV, in the
following referred to as \esi , the dispersing element
D$_{\indrm{\an}}$ should feature $|\fcomm{\an}/\dcomm{\an}| \simeq
25~\mu$rad/meV.  For practical reasons, we assume here that the
secondary {\em monochromatic} source size is $\Delta x_{\ind{1}}\simeq
5~\mu$m, which is presently routinely achievable, and the focal length
of the collimating element in the refocusing system is
$f_{\ind{1}}\simeq 0.2$~m, the value which ensures collection of
x~rays scattered from the sample in a large solid angle. An x-ray echo
spectrometer with a resolution $\Delta \etra = 0.1$~meV, hereafter
referred to as \esoi , requires also a better momentum
transfer resolution, i.e., a smaller solid angle of collection.
Assuming, therefore, a larger focal distance $f_{\ind{1}}\simeq 0.4$~m,
we obtain $|\fcomm{\an}/\dcomm{\an}|\simeq 125~\mu$rad/meV in this
case.  In the following, we will gradually specify  parameters of
the exemplary echo-type IXS spectrometers \esi\ and \esoi, and
list them in Table~\ref{tab-xes}.

Now, with the $|\fcomm{\an}/\dcomm{\an}|$ and $f_{\ind{1}}$ values
being specified, Eq.  \eqref{refocus23} can be used to estimate the
required cumulative dispersing rate ${|\fcomm{\mo}|}$ of the
dispersing element D$_{\indrm{\mo}}$.  Assuming a comfortable distance
$l_{\ind{3}}\simeq 2$~m from the focusing element F to the sample in the
defocusing system, we estimate ${|\fcomm{\mo}|} \simeq
2.5~\mu$rad/meV for spectrometer \esi\ and ${|\fcomm{\mo}|} \simeq
25~\mu$rad/meV for  \esoi , respectively.

\section{Dispersive optic}
\label{doptic}

In the next step, we consider optical designs of the dispersing
elements in the hard x-ray regime which could deliver the required
values of the angular dispersion rates discussed in the previous
section.

\begin{figure}[t!]
\includegraphics[width=0.5\textwidth]{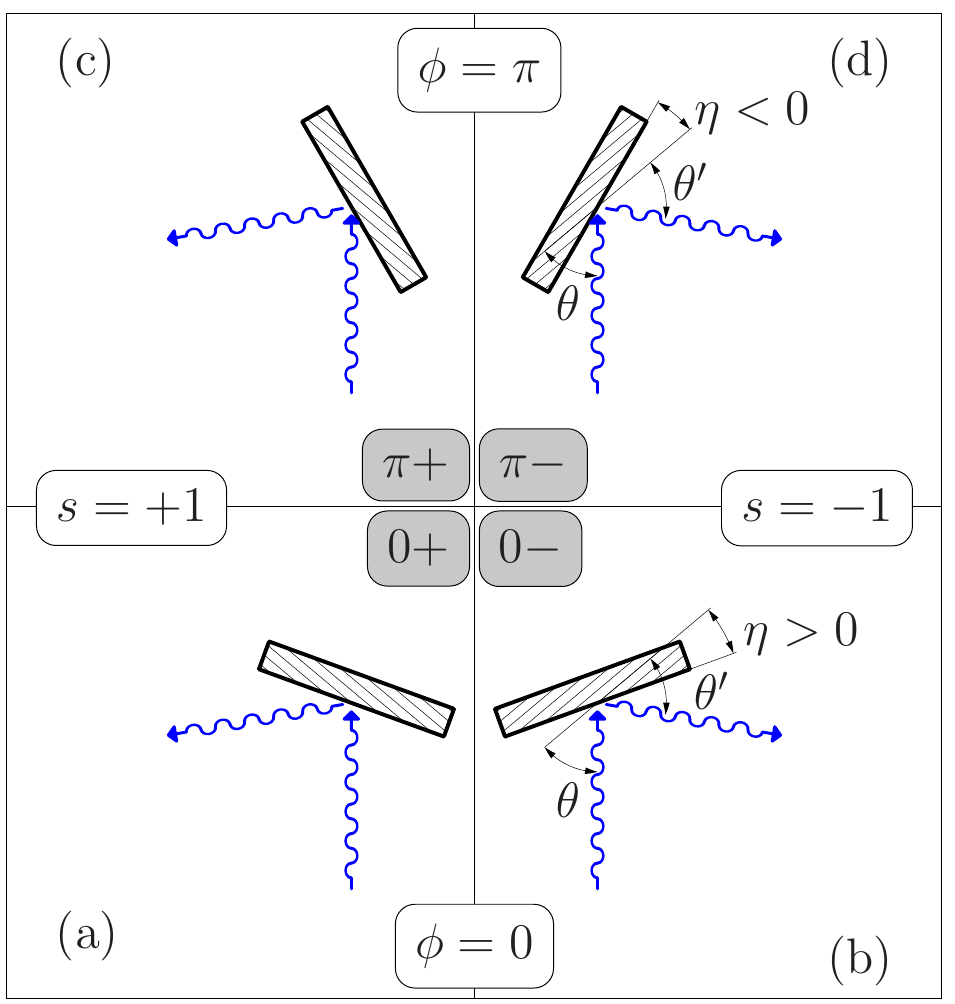}
\caption{Definition of scattering geometries in Bragg diffraction from
  asymmetrically cut crystals, with asymmetry angle $\eta$. Either
  geometry is specified by two parameters $\phi$ and $\sgn$, each
  taking two possible values $\phi=0,\pi$ and $s=\pm1$ : (a) $0+$, (b)
  $0-$, (c) $\pi+$, and (d) $\pi-$. The ``deflection'' sign $\sgn=+1$
  corresponds to reflection in the counterclockwise direction as in
  (a) and (c), while $\sgn=-1$ means the clockwise direction as in (b)
  and (d). The azimuthal angle of incidence $\phi=0$ (see definition in
  \cite{Shvydko-SB}) is equivalent here to a positive asymmetry angle
  $\eta$
as in (a) and (b).  The reversed scattering geometries shown in (c) and (d)
correspond to $\phi=\pi$ and equivalently $\eta<0$.}
\label{fig005}
\end{figure}

\subsection{Angular dispersion rate}  
\label{angulardispersionrate}  

The angular dispersion rate $\dirate={{\mathrm
    d}\theta^{\prime}}/{{\mathrm d}E} $ measures the variation with
photon energy $E$ of the glancing angle of
reflection $\theta^{\prime}$  from the Bragg diffracting atomic planes, assuming the glancing
angle of incidence $\theta$ (Bragg angle) is fixed.  The angular
dispersion rate \cite{MK80-2,Shvydko-SB,SSM13}
\begin{equation}                                                                                                                    
\dirate=\frac{2\sin\theta \sin{\eta}}{E\sin(\theta^{\prime}-\eta)} \equiv -\frac{1}{E}(1+b)\tan\theta.       
\label{ad010}
\end{equation}
is nonzero only if the ``asymmetry'' angle $\eta$ between the atomic
planes and the crystal surface is nonzero. Here
\begin{equation}                                                                                                             
b=-\frac{\sin(\theta+\eta)}{\sin(\theta^{\prime}-\eta)}
\label{ad020}
\end{equation}
is the asymmetry ratio. The angle $\eta$ and its sign are defined in
Fig.~\ref{fig005}.

The dispersion rate is biggest, first, in Bragg backscattering when
$\theta \rightarrow \pi/2$; second, when $\theta^{\prime}-\eta
\rightarrow 0$, i.e.,  when x~rays are reflected at grazing emergence
to the crystal surface as in Figs.~\ref{fig005}(a)-(b); and, third,
for x~rays with smaller photon energies. In the following examples we
use the (008) Bragg back reflection from Si of x~rays with photon
energy $E\simeq 9.1$~keV. Such energy is optimal, ensuring sufficiently
large dispersion rate and yet not too large
photoabsorption in the optical elements and the sample.

The variation  ${\mathrm d}\theta^{\prime}$ and the difference
$|\theta-\theta^{\prime}|\lesssim 10^{-5}$ are very small, and
therefore in most cases $\theta^{\prime}$ in
Eqs.~\eqref{ad010}-\eqref{ad020} can be replaced by $\theta$.

The cumulative dispersion rate $\fcomm{n}$ of a system of sequentially
diffracting $n$ crystals can be calculated using the
recursive relationship \cite{SSM13,Shvydko15}
\begin{equation}
\fcomm{n}=b_{\ind{n}}\fcomm{n-1} + \sgn_{\ind{n}}\dirate_{\ind{n}},
\label{ad030}
\end{equation}
with the deflection signs $\sgn_{\ind{n}}=\pm 1$ defined in
Fig.~\ref{fig005}.  Remarkably, if the asymmetry ratio of the last
$n$-th crystal is large $|b_{\ind{n}}|\gg 1$, which can take place if
$\eta>0$ ($\phi=0$) as in Figs.~\ref{fig005}(a)-(b), the cumulative
dispersion rate $\fcomm{n-1}$ of the previous $n-1$ crystals can be
amplified significantly, resulting in a very large cumulative
dispersion rate $\fcomm{n}$ of the whole system \cite{SSM13}.

\subsection{One-crystal dispersing elements}  

\begin{figure}
\includegraphics[width=0.5\textwidth]{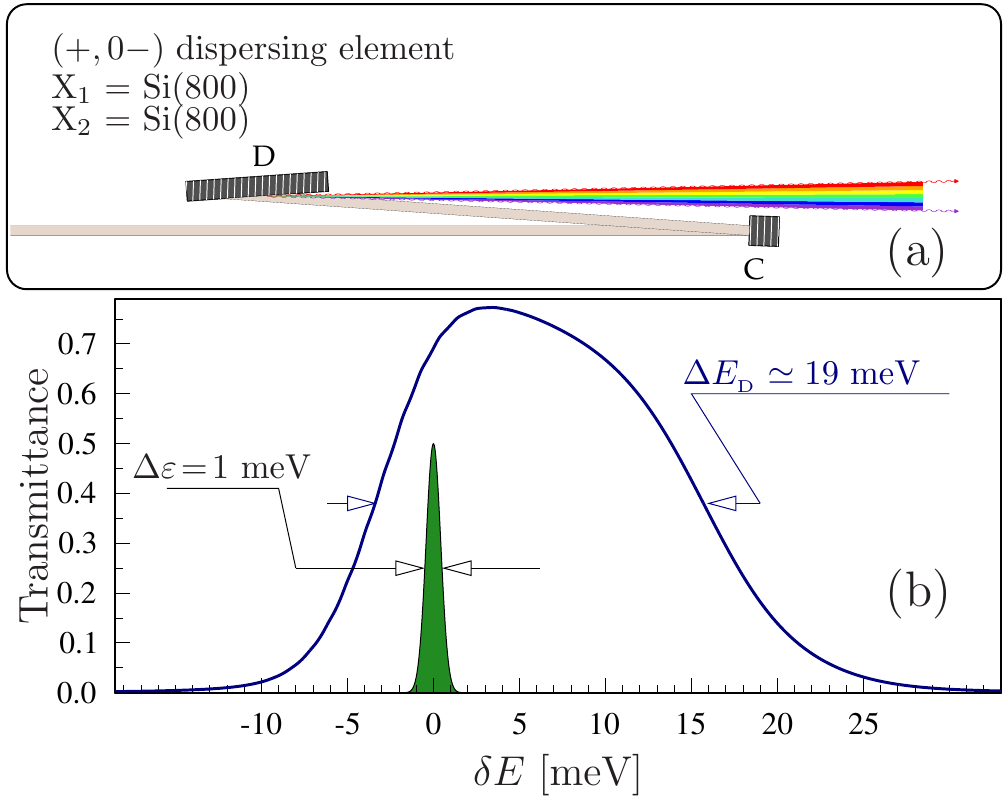}
\caption{X-ray dispersing element composed of one asymmetrically cut
  crystal D (a) and its spectral transmittance function (b) calculated
  for the incident beam divergence of $20~\mu$rad.  The symmetrically
  cut crystal C is added to ensure the in-line scheme. With the
  crystal parameters provided in Table~\ref{tab-1meV},
the dispersing element features a spectral transmission function with
a $\dband=19$~meV bandwidth (b), a cumulative angular dispersion rate
$\fcomm{\mo} = -3.12 ~\mu$rad/meV, and a cumulative asymmetry factor
$\dcomm{\mo} = 2.0$, appropriate for dispersing element
D$_{\indrm{\mo}}$ of the defocusing system $\hat{O}_{\indrm{\mo}}$
(see Fig.~\ref{fig002}) of the 1-meV-resolution x-ray echo
spectrometer \esi .  The sharp green line in (b) indicates the 1-meV
design spectral resolution. }
\label{fig0012}
\end{figure}

The simplest x-ray dispersing element consists of one asymmetrically cut
crystal.  The largest attainable dispersion rate in Bragg diffraction
of $\simeq 9$~keV x~rays from one crystal is $\dirate\lesssim
10~\mu$rad/meV. This follows from Eq.~\eqref{ad010} by applying
extreme but yet realistic values for $\theta\simeq
88^{\circ}$-$89^{\circ} $ and $\theta-\eta \gtrsim 1^{\circ}$. A
one-crystal dispersing element is applicable if the required
dispersion rate is smaller. This is the case of the dispersing element
D$_{\indrm{\mo}}$ of the defocusing system of the 1-meV-resolution
spectrometer \esi\ requiring ${|\fcomm{\mo}|} \simeq 2.5~\mu$rad/meV.
Figure~\ref{fig0012} shows an example of an optical design and
spectral transmission function of the dispersing element. The function
of the  additional {\em symmetrically} cut ($\eta=0$) crystal C
is merely to keep the dispersed beam average direction after
reflection from the asymmetrically cut crystal D parallel
to the direction of the incident beam (in-line scheme).
%

\begin{figure}
\includegraphics[width=0.5\textwidth]{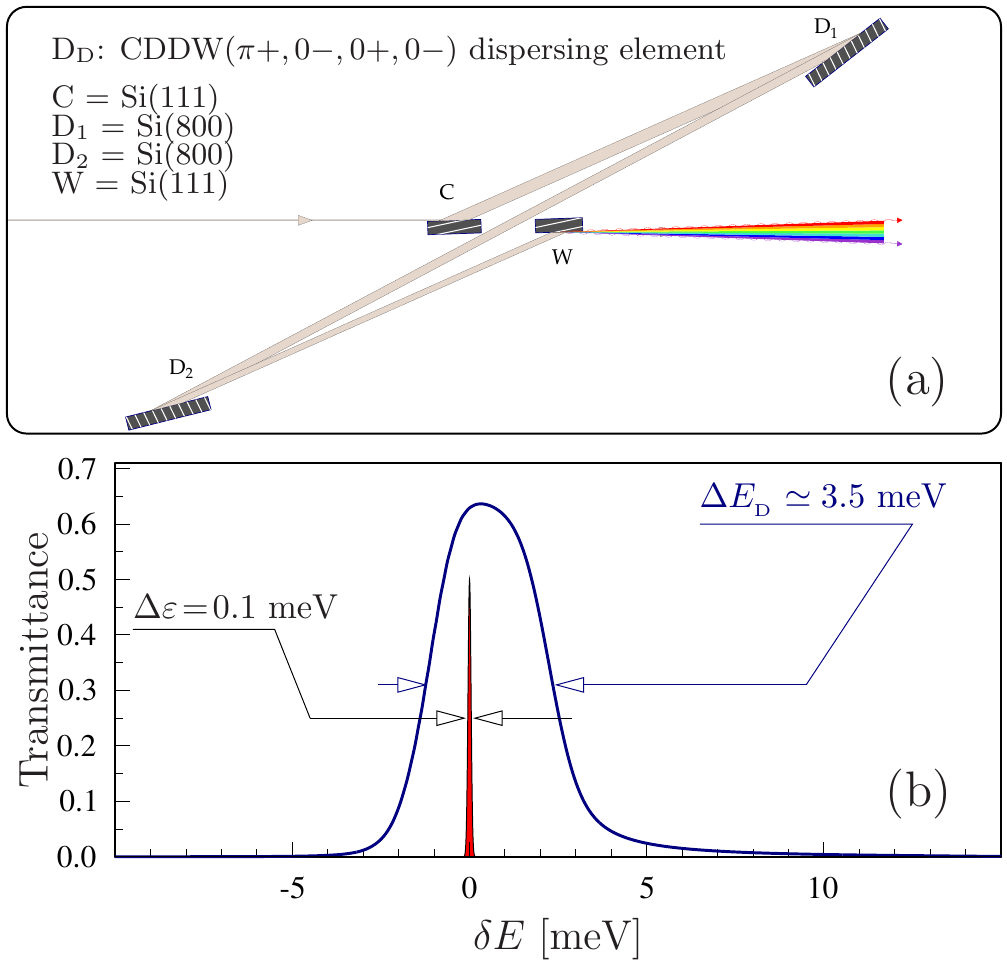}
  \caption{In-line four-crystal CDDW-type x-ray dispersing element in
    a ($\pi+$,$0-$,$0+$,$0-$) scattering configuration (a), and its
    spectral transmittance function (b) calculated for the incident
    beam divergence of $20~\mu$rad.  With the crystal parameters
    provided in Table~\ref{tab-0o1meV},
the optic features a spectral transmission function with a
$\dband=3.5$~meV bandwidth (b), a cumulative angular dispersion rate
$\fcomm{\mo} = -32 ~\mu$rad/meV, and a cumulative asymmetry factor
$\dcomm{\mo} = 2$ appropriate for dispersing element D$_{\indrm{\mo}}$
of the defocusing system $\hat{O}_{\indrm{\mo}}$ (see
Fig.~\ref{fig002}) of the 0.1-meV-resolution x-ray echo spectrometer
\esoi .  The sharp red line in (b) indicates the 0.1-meV design
spectral resolution. }
\label{fig003}
\end{figure}

\begin{figure}
\includegraphics[width=0.5\textwidth]{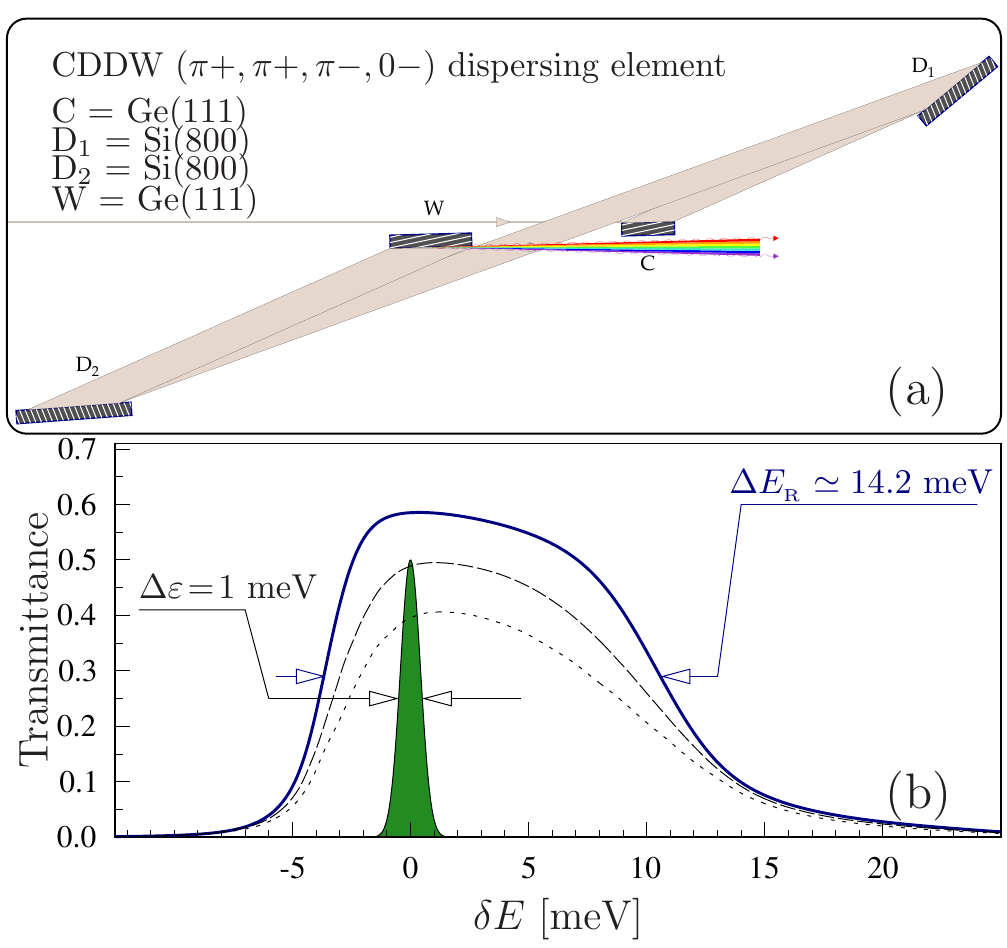}
\caption{In-line four-crystal CDDW-type x-ray dispersing element
  similar to that in Fig.~\ref{fig003} but in a
  ($\pi+$,$\pi+$,$\pi-$,$0-$) scattering configuration (a), and its
  spectral transmittance function (b) calculated for the incident beam
  divergence of $100~\mu$rad (bold), $200~\mu$rad (dashed), and
  $300~\mu$rad (dotted). With the crystal parameters provided in
  Table~\ref{tab-1meV},
the optic features a $\rband=14.2$~meV bandwidth (b), a cumulative
angular dispersion rate $\fcomm{\an} = -16.47 ~\mu$rad/meV, a
cumulative asymmetry factor $\dcomm{\an} = 0.65$, and
$\fcomm{\an}/\dcomm{\an} = -25.06 ~\mu$rad/meV, appropriate for
dispersing element D$_{\indrm{\an}}$ of the refocusing system
$\hat{O}_{\indrm{\an}}$ (see Fig.~\ref{fig002}) of the
1-meV-resolution x-ray echo spectrometer \esi .  The sharp line in (b)
presents the 1-meV design spectral resolution $\Delta \epsilon$ of the
x-ray echo spectrometer.}
\label{fig004v4}
\end{figure}

\begin{table}
\centering
\begin{tabular}{|l|lllllll|}
  \hline   \hline 
crystal   & $\vc{H}_{\indrm{\elmt}}$ &$\eta_{\indrm{\elmt}} $ &$\theta_{\indrm{\elmt}} $  & $\deis{\elmt} $ &  $\dais{\elmt}$  & $b_{\indrm{\elmt}}$ & $\sgn_{\indrm{\elmt}}\dirate_{\indrm{\elmt}} $ \\[-5pt]    
element (\elmt)  & &  &  &   &  &   &    \\[0pt]    
[material]          & $(hkl)$ & deg & deg  & meV  &  $\mu$rad   & & $\frac{\mu {\mathrm {rad}}}{\mathrm {meV}}$ \\[5pt]    
\hline  \hline  
\multicolumn{8}{|l|}{D$_{\indrm{\mo}}$: ($+$,$0-$), Fig.~\ref{fig0012}}\\  
\cline{1-1}
1~~C~[Si] & (8~0~0) &  0   &  88   & 27 &  85    & -1.0  & 0 \\[-0.0pt]
2~~D~[Si] & (8~0~0) &  84  &  88   & 27 &  85   & -2.0  & -3.12 \\[-0.0pt]
  \hline  
\multicolumn{3}{|l}{Cumulative values}  & $\daccept$ & $\dband$  & $\ddtheta$  & $\dcomm{\mo}$ & $\fcomm{\mo}$ \\
\multicolumn{3}{|l}{} & $\mu$rad   & meV  & $\mu$rad  &  & $\frac{\mu {\mathrm {rad}}}{\mathrm {meV}}$ \\[5pt]    
\cline{4-8}
\multicolumn{3}{|l}{} & 59 & 19  & 59.6  & 2.0 & -3.12 \\
\hline  \hline  
\multicolumn{8}{|l|}{D$_{\indrm{\an}}$: CDDW ($\pi+$,$\pi+$,$\pi-$,$0-$), Fig.~\ref{fig004v4}}\\  
\cline{1-1}  
1~~C~~[Ge]& (1~1~1) &  -10.5  &  12.0  &  3013 & 71    & -0.069  & -0.022 \\[-0.0pt]
2~~D$_{\ind{1}}$~[Si] & (8~0~0) &  -72.2  &  88  &  27 &  85    & -0.80  & -0.62 \\[-0.0pt]
3~~D$_{\ind{2}}$~[Si] & (8~0~0) &  -72.2  &  88   & 27 &  85   & -0.80  & +0.62 \\[-0.0pt]
4~~W~~[Ge] & (1~1~1) &  10.5  &  12.0   & 3013 &  71   & -14.8  & -0.31 \\[-0.0pt]
  \hline  
\multicolumn{3}{|l}{Cumulative values} & $\raccept$ & $\rband$  & $\rdtheta$  & $\dcomm{\an}$ & $\fcomm{\an}$ \\
\multicolumn{3}{|l}{} & $\mu$rad & meV  & $\mu$rad  &  & $\frac{\mu {\mathrm {rad}}}{\mathrm {meV}}$ \\[5pt]    
\cline{4-8}
\multicolumn{3}{|l}{} &246   & 14.2  & 234  & 0.65 & -16.5
 \\
  \hline  
  \hline  
\end{tabular}
\caption{Examples of in-line crystal optics as dispersing elements
  (``diffraction gratings'') D$_{\indrm{D}}$, D$_{\indrm{R}}$ of the
  defocusing $\hat{O}_{\indrm{D}}$ and refocusing
  $\hat{O}_{\indrm{\an}}$ systems of the 1-meV-resolution x-ray echo
  spectrometer \esi .  For each optic, the table presents crystal
  elements (\elmt=C,D$_{\ind{1}}$,D$_{\ind{2}}$,W) and their Bragg
  reflection parameters: $(hkl)$, Miller indices of the Bragg
  diffraction vector $\vc{H}_{\indrm{\elmt}}$; $\eta_{\indrm{\elmt}}$,
  asymmetry angle; $\theta_{\indrm{\elmt}}$, glancing angle of
  incidence; $\deis{\elmt}$, $\dais{\elmt}$ Bragg reflection intrinsic
  spectral width and angular acceptance in symmetric scattering
  geometry, respectively; $b_{\indrm{\elmt}}$, asymmetry ratio; and
  $\sgn_{\indrm{\elmt}} \dirate_{\indrm{\elmt}} $, angular dispersion
  rate with deflection sign.  For each optic, also shown are: angular
  acceptance $\xaccept$ (X=D,R) and spectral bandwidth $\xband$ as
  derived from the dynamical theory calculations, the angular spread
  of the dispersion fan $\xdtheta=|\fcomm{X}| \xband$, and the
  cumulative values of the asymmetry parameter $\dcomm{X}$ and the
  dispersion rate $\fcomm{X}$.  X-ray photon energy is
  $E=9.13708$~keV.}
\label{tab-1meV}
\end{table}

\begin{table}
\centering
\begin{tabular}{|l|lllllll|}
  \hline   \hline 
crystal   & $\vc{H}_{\indrm{\elmt}}$ &$\eta_{\indrm{\elmt}} $ &$\theta_{\indrm{\elmt}} $  & $\deis{\elmt} $ &  $\dais{\elmt}$  & $b_{\indrm{\elmt}}$ & $\sgn_{\indrm{\elmt}}\dirate_{\indrm{\elmt}} $ \\[-5pt]    
element (\elmt)  & &  &  &   &  &   &    \\[0pt]    
[material]  & $(hkl)$ & deg & deg  & meV  &  $\mu$rad   & & $\frac{\mu {\mathrm {rad}}}{\mathrm {meV}}$ \\[5pt]    
\hline  \hline  
\multicolumn{8}{|l|}{D$_{\indrm{\mo}}$: CDDW ($\pi+$,$0-$,$0+$,$0-$), Fig.~\ref{fig003}}\\  
\cline{1-1}
1~~C~~[Si]& (1~1~1) &  -10.5  &  12.5  &  1304 & 32    & -0.09  & -0.02 \\[-0.0pt]
2~~D$_{\ind{1}}$~[Si] & (8~0~0) &  77.7  &  88  &  27 &  85    & -1.38  & -1.19 \\[-0.0pt]
3~~D$_{\ind{2}}$~[Si] & (8~0~0) &  77.7  &  88   & 27 &  85   & -1.38  & +1.19 \\[-0.0pt]
4~~W~~[Si] & (1~1~1) &  10.5  &  12.5   & 3013 &  71   & -11.2  & -0.24 \\[-0.0pt]
  \hline  
\multicolumn{3}{|l}{Cumulative values} & $\daccept$  & $\dband$  & $\ddtheta$  & $\dcomm{\mo}$ & $\fcomm{\mo}$ \\
\multicolumn{3}{|l}{} & $\mu$rad   & meV  & $\mu$rad  &  & $\frac{\mu {\mathrm {rad}}}{\mathrm {meV}}$ \\[5pt]    
\cline{4-8}
\multicolumn{3}{|l}{} & 57 & 3.5  & 112  & 1.91 & -31.7 \\
\hline  \hline  
\multicolumn{8}{|l|}{D$_{\indrm{\an}}$: CDDW ($\pi+$,$\pi+$,$\pi-$,$0-$), Fig.~\ref{fig004v3}}\\  
\cline{1-1}
1~~C~~[Ge]& (1~1~1) &  -10.5  &  12.0  &  3013 & 71    & -0.07  & -0.02 \\[-0.0pt]
2~~D$_{\ind{1}}$~[Si] & (8~0~0) &  -83.75  &  88  &  27 &  85    & -0.52  & -1.50 \\[-0.0pt]
3~~D$_{\ind{2}}$~[Si] & (8~0~0) &  -83.75  &  88   & 27 &  85   & -0.52  & +1.50 \\[-0.0pt]
4~~W~~[Ge] & (1~1~1) &  10.5  &  12.0   & 3013 &  71   & -14.75  & -0.31 \\[-0.0pt]
  \hline  
\multicolumn{3}{|l}{Cumulative values} & $\raccept$ & $\rband$  & $\rdtheta$  & $\dcomm{\an}$ & $\fcomm{\an}$ \\
\multicolumn{3}{|l}{} & $\mu$rad  & meV  & $\mu$rad  &  & $\frac{\mu {\mathrm {rad}}}{\mathrm {meV}}$ \\[5pt]    
\cline{4-8}
\multicolumn{3}{|l}{} & 262  & 8  & 272  & 0.27 & -34.15 \\
  \hline  
  \hline  
\end{tabular}
\caption{Same as Table~\ref{tab-1meV}, but for  the 0.1-meV-resolution x-ray echo
  spectrometer \esoi .}
\label{tab-0o1meV}
\end{table}

\begin{figure}
\includegraphics[width=0.5\textwidth]{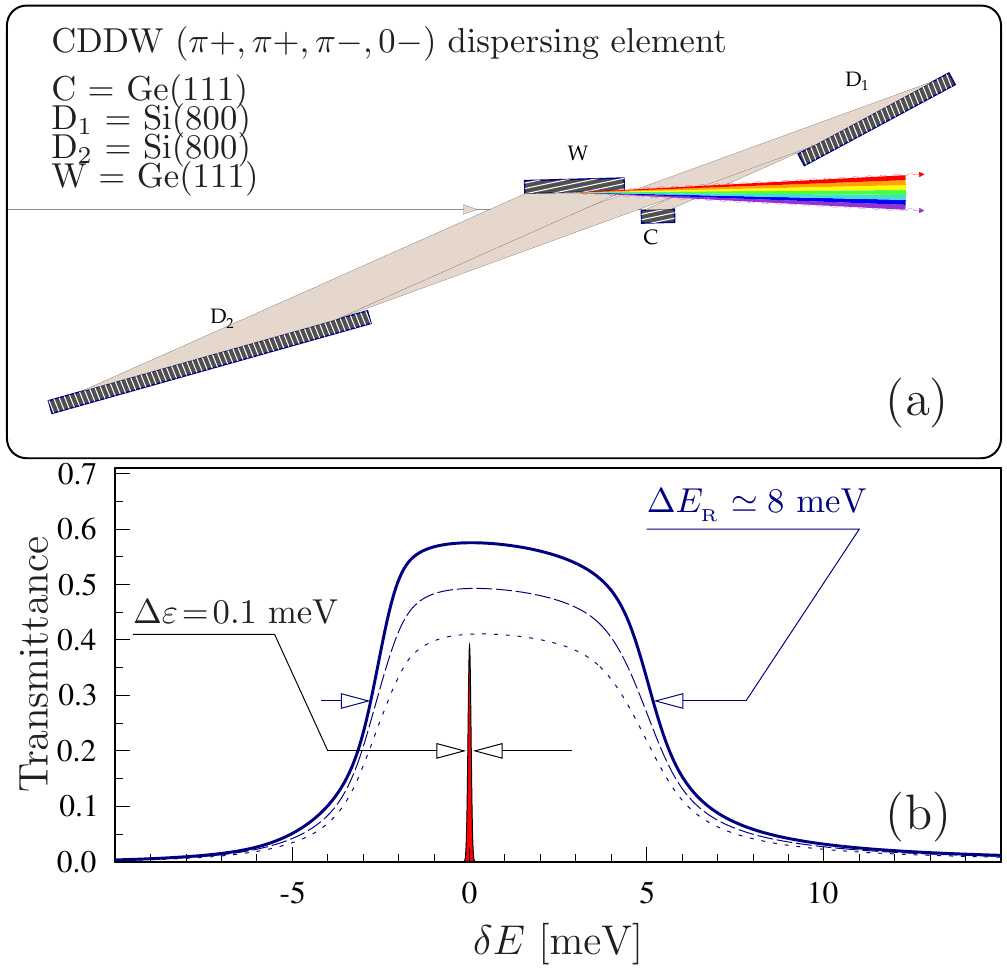}
\caption{In-line four-crystal CDDW-type x-ray dispersing element in a
  ($\pi+$,$\pi+$,$\pi-$,$0-$) scattering configuration (a), and its
  spectral transmittance function (b) calculated for the incident beam
  divergence of $100~\mu$rad (bold), $200~\mu$rad (dashed), and
  $300~\mu$rad (dotted).  With the crystal parameters provided in
  Table~\ref{tab-0o1meV}, the optic features a $\rband=8$~meV
  bandwidth (b), a cumulative angular dispersion rate $\fcomm{\an} =
  -34.2 ~\mu$rad/meV, a cumulative asymmetry factor $\dcomm{\an} =
  0.27$, and $\fcomm{\an}/\dcomm{\an} = -125.5 ~\mu$rad/meV,
  appropriate for dispersing element D$_{\indrm{\an}}$ of the
  refocusing system $\hat{O}_{\indrm{\an}}$ (see Fig.~\ref{fig002}) of
  the 0.1-meV-resolution x-ray echo spectrometer \esoi .  The sharp
  line in (b) presents the 0.1-meV design spectral resolution $\Delta
  \epsilon$ of the x-ray echo spectrometer.}
\label{fig004v3}
\end{figure}

\subsection{Four-crystal CDDW dispersing elements}  
\label{cddw}

Dispersion elements with dispersion rates more than $\simeq
10~\mu$rad/meV require multicrystal solutions, ensuring dispersion
rate enhancement according to Eq.~\eqref{ad030}.  In \cite{SSM13} it
was demonstrated that the angular dispersion rate of a four-crystal CDDW
optic \cite{ShSS11,SSS13,SSS14}, schematically shown in
Figs.~\ref{fig003}, \ref{fig004v4}, and \ref{fig004v3}, can be
dramatically enhanced by almost two orders of magnitude compared to
what is possible with one asymmetrically cut crystal.  The CDDW optic
is not unique in achieving large dispersion rates. But, as discussed
further in more detail, the CDDW optic is advantageous, as it features
also a large angular acceptance, especially valuable for the
refocusing dispersing element, and relatively large spectral
bandwidths.  The CDDW-type dispersing optics are therefore proposed
here for use as large-dispersion-rate dispersing elements.

The in-line four-crystal CDDW-type dispersing optic comprises
collimating (C), dispersing (D$_{\ind{1}}$, D$_{\ind{2}}$), and
wavelength-selecting (W) crystals, which can be arranged in different
scattering configurations.  In the general case, a four-crystal
scattering configuration can be defined as
$(\phi_{\ind{1}}\sgn_{\ind{1}},\phi_{\ind{2}}\sgn_{\ind{2}},\phi_{\ind{3}}\sgn_{\ind{3}},\phi_{\ind{4}}\sgn_{\ind{4}})$.
Here, for each crystal (C=1, D$_{\ind{1}}=2$, D$_{\ind{2}}=3$, W=4)
the $\phi_{\ind{n}}$ and $\sgn_{\ind{n}}$ values ($n=1,2,3,4$) define
the scattering geometry on each crystal, as in
Fig.~\ref{fig005}. Without loss of generality, we set for distinctness
in all cases $\sgn_{\ind{1}}=+1$. To ensure a large angular acceptance
and collimation, which is possible if $|b_{\ind{1}}|\ll 1$ is chosen
for the first crystal, we set $\phi_{\ind{1}}=\pi$.  To ensure large
dispersion rate enhancement, a large $|b_{\ind{4}}|$ is
needed. Therefore, we set $\phi_{\ind{4}}=0$.  Of all the rest of the
32 possible cases
$(\pi+,\phi_{\ind{2}}\sgn_{\ind{2}},\phi_{\ind{3}}\sgn_{\ind{3}},0\sgn_{\ind{4}})$,
those scattering geometries will be considered which feature an
in-line scheme, the largest cumulative dispersion rate $|\fcomm{n}|$
for the dispersing element of the defocusing system, and the largest
$|\fcomm{n}/\dcomm{n}|$ value in case of the dispersing element of the
refocusing system.

Following Eq.~\eqref{ad030}, the cumulative dispersion rate
$\fcomm{4}$ in a four-crystal system is given in the general case by
\begin{equation}
\fcomm{4} =
b_{\ind{4}}b_{\ind{3}}b_{\ind{2}}\sgn_{\ind{1}}\dirate_{\ind{1}}+b_{\ind{4}}b_{\ind{3}}\sgn_{\ind{2}}\dirate_{\ind{2}}+b_{\ind{4}}\sgn_{\ind{3}}\dirate_{\ind{3}}+\sgn_{\ind{4}}\dirate_{\ind{4}}.
\label{scum010}
\end{equation}
Low-index Bragg reflections with small Bragg angles are typically
chosen for the C and W~crystals ($n=1,4$) to ensure large angular
acceptance and broadband transmission functions. On the contrary,
high-index Bragg reflections with Bragg angles close to $90^{\circ}$
are chosen to ensure the large dispersion rates of the D~crystals
($n=2,3$), which are typically much larger than those of the C and
W~crystals. Under theses conditions, the expression for the cumulative
dispersion rate can be reduced to $\fcomm{4} \simeq
b_{\ind{4}}b_{\ind{3}}\sgn_{\ind{2}}\dirate_{\ind{2}}+b_{\ind{4}}\sgn_{\ind{3}}\dirate_{\ind{3}}=b_{\ind{4}}(b_{\ind{3}}\sgn_{\ind{2}}\dirate_{\ind{2}}+\sgn_{\ind{3}}\dirate_{\ind{3}})$.
Since $b_{\ind{3}} < 0$, the largest dispersion rates can be achieved
in the systems for which the product
$\sgn_{\ind{2}}\sgn_{\ind{3}}\dirate_{\ind{2}}\dirate_{\ind{3}}<0$. In
this case, and assuming $|\dirate_{\ind{2}}|=|\dirate_{\ind{3}}|$, we obtain
\begin{equation}
\fcomm{4} \simeq b_{\ind{4}}\sgn_{\ind{3}}\dirate_{\ind{3}}(1-b_{\ind{3}}).
\label{scum020}
\end{equation}
Optical designs with $b_{\ind{4}}\simeq -20$ and
$b_{\ind{2}}=b_{\ind{3}}\simeq -4$ may ensure enhancement of the
cumulative dispersion rate of up to two orders of magnitude compared
to what is possible with one crystal.

There are four large-dispersion-rate CDDW configurations featuring
$\dirate_{\ind{2}}\dirate_{\ind{3}}<0$ and
$\sgn_{\ind{2}}\sgn_{\ind{3}}>0$: $(\pi+,\pi-,0-,0-)$;
$(\pi+,\pi+,0+,0-)$; $(\pi+,0-,\pi-,0-)$; and $(\pi+,0+,\pi+,0-)$.
However, the angle between the incident and reflected beams is
$4(\theta_{\indrm{2}}-\pi/2)$, i.e. the beams impinging upon and
emerging from the system are not parallel.

There are four other  large-dispersion-rate CDDW configurations
featuring $\dirate_{\ind{2}}\dirate_{\ind{3}}>0$ and
$\sgn_{\ind{2}}\sgn_{\ind{3}}<0$: $(\pi+,\pi-,\pi+,0-)$;
$(\pi+,\pi+,\pi-,0-)$; $(\pi+,0-,0+,0-)$; and $(\pi+,0+,0-,0-)$. These
configurations are especially interesting, because the incident and
transmitted x~rays are parallel (in-line scheme).

In the present paper, we choose the in-line large-dispersion-rate CDDW
optic in the ($\pi+$,$0-$,$0+$,$0-$) configuration, with
$|b_{\ind{2}}|=|b_{\ind{3}}| > 1$; see example in Fig.~\ref{fig003}
for the dispersing elements D$_{\indrm{\mo}}$ of the defocusing
systems. This configuration is preferred as it provides large
dispersion rates $|\fcomm{\mo}|$ [see Eq.~\eqref{scum020}] significant
transmission bandwidth $\dband$, and compactness.

The CDDW optic in the ($\pi+$,$\pi+$,$\pi-$,$0-$) configuration, with
$|b_{\ind{2}}|=|b_{\ind{3}}| < 1$ is better suited for the refocusing
dispersing elements D$_{\indrm{\an}}$ (Figs.~\ref{fig004v4} and \ref{fig004v3}). It provides
large absolute values of the ratio
\begin{equation}
\frac{\fcomm{4}}{\dcomm{4}} \simeq -\sgn_{\ind{3}}\dirate_{\ind{3}}\frac{1-b_{\ind{3}}}{b_{\ind{1}}b_{\ind{2}}b_{\ind{3}}}.
\label{scum030}
\end{equation}
required for the high spectral resolution of the echo spectrometers
[see Eq.~\eqref{resolution2}], substantial transmission bandwidths
$\rband/ \Delta \etra \gg 1$, and large angular acceptance values
$\raccept \simeq 250~\mu$rad; see Tables~\ref{tab-1meV} and
\ref{tab-0o1meV}.

Examples of the dispersing elements and their crystal parameters for
the 1-meV-resolution x-ray echo spectrometer \esi\ are provided in
Figs.~\ref{fig0012} and \ref{fig004v4} and Table~\ref{tab-1meV}.  For
the 0.1-meV-resolution x-ray echo spectrometer \esoi , they are
provided in Figs.~\ref{fig003} and ~\ref{fig004v3} and 
Table~\ref{tab-0o1meV}.

\section{Effective vertical beam size on the sample and effective spectral bandwidth}
\label{vspotsize} 

\begin{figure}
\includegraphics[width=0.5\textwidth]{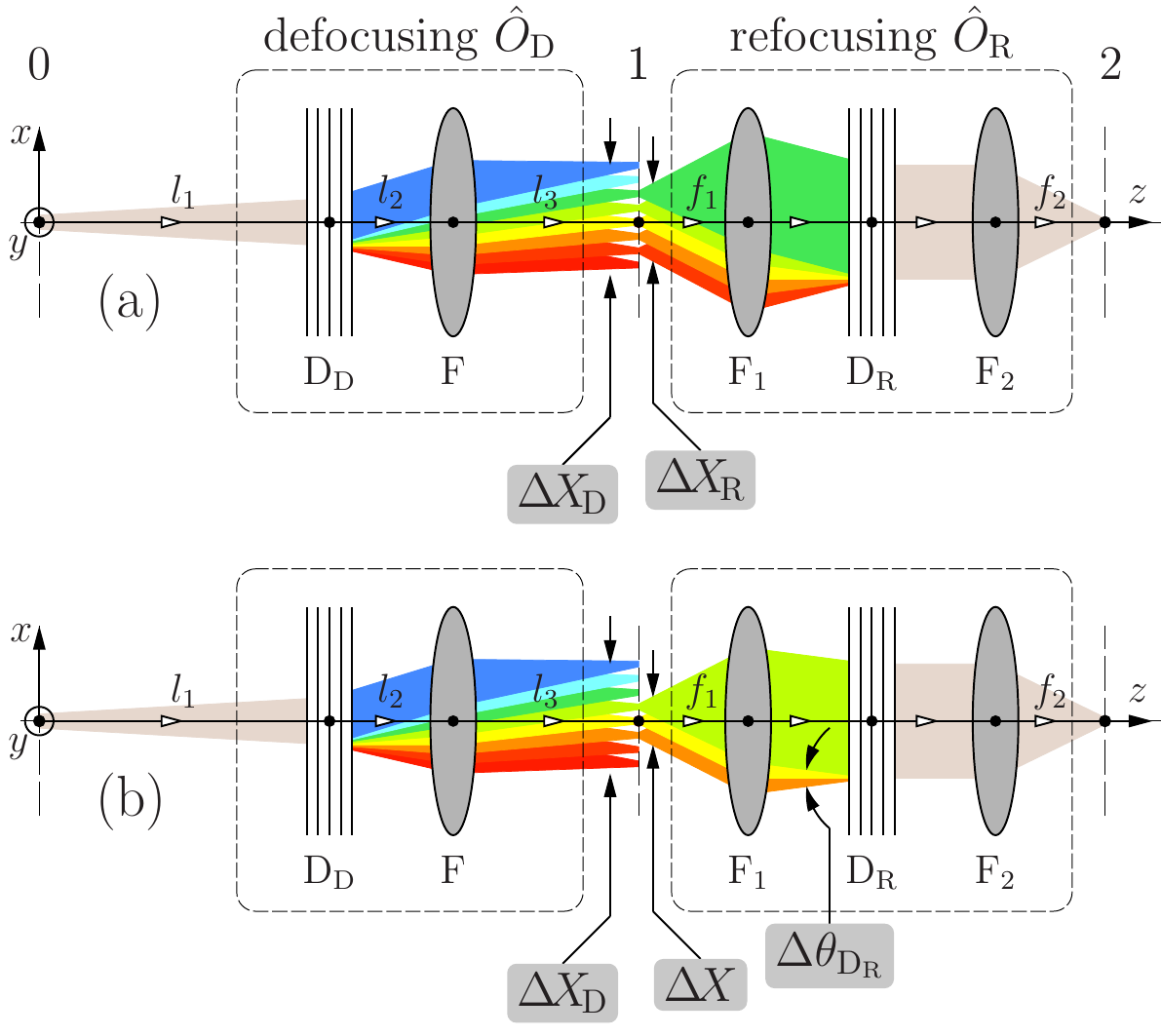}
\caption{Optical scheme of an x-ray echo spectrometer in the vertical
  scattering (dispersion) plane detailing the vertical beam size $\Delta
  X_{\indrm{D}}$ on the sample and the reduced effective vertical beam
  sizes $\Delta X_{\indrm{R}}$ and $\Xsize$ as
  seen by the refocusing system due to a smaller spectral bandwidth
  (a) or the limited angular acceptance $ \Delta \theta_{\indrm{D_R}} $ (b). }
\label{fig00vsp}
\end{figure}

Each monochromatic spectral component is focused onto the sample in
reference plane $1$ to a spot with a vertical size of $\Delta
x_{\ind{1}}$. However, different spectral components are focused at
different positions distributed over a length of
\begin{equation}
\Delta X_{\indrm{D}} = |G_{\indrm{D}}| \dband 
\label{vsp010} 
\end{equation}
on the sample; see Fig.~\ref{fig00vsp}. Here $\dband $ is the total
spectral width of x~rays incident on the sample. In the limit 
$\Delta x_{\ind{1}} \ll \Delta X_{\indrm{D}}$, which is considered
here, the vertical beam size on the sample in the dispersion plane $(x,z)$ is $\Delta X_{\indrm{D}} $.

The effective vertical beam size $\Delta X_{\indrm{R}}$ as seen by the
refocusing system may differ from $\Delta X_{\indrm{D}}$.
Particularly, this happens if the spectral bandwidth $\rband$ of the
refocusing system is smaller than $\dband $.  In this case, the
effective beam size $\Delta X_{\indrm{R}}$ is smaller than $\Delta
X_{\indrm{D}}$ [see Fig.~\ref{fig00vsp}(a)] and is given by
\begin{equation} 
\Delta X_{\indrm{R}} = |G_{\indrm{D}}| \rband  \equiv \left| \frac{\fcomm{\an}}{\dcomm{\an}}\right| \, f_{\ind{1}} \,\rband .
\label{vsp020} 
\end{equation}
The right-hand side of Eq.~\eqref{vsp020} is derived from the refocusing
condition Eq.~\eqref{refocus} and Eq.~\eqref{refocusing}. Further, applying Eq.~\eqref{resolution2},
the effective beam size $\Delta X_{\indrm{R}} $ 
can be presented in an equivalent 
form
\begin{equation} 
\Delta X_{\indrm{R}} = \Delta x_{\ind{1}} \frac{\rband }{\Delta\varepsilon} ,   
\label{vsp025} 
\end{equation}
expressed through the required spectral resolution
$\Delta\varepsilon$ of the spectrometer and the secondary source size
$\Delta x_{\ind{1}}$.

The effective beam size on the sample can become even smaller.  Indeed, if
the angular acceptance $\raccept $ of the
dispersing element D$_{\indrm{\an}}$ is smaller than the effective
angular spread $\simeq \Delta X_{\indrm{R}}/f_{\ind{1}}$ of the beam incident
on D$_{\indrm{\an}}$, then the effective beam size accepted by the refocusing system is further
reduced to
\begin{equation}
\Xsize \simeq f_{\ind{1}} \raccept ;
\label{vsp030} 
\end{equation}
see Fig.~\ref{fig00vsp}(b). 

By the same reasoning, the spectral bandwidth of the incident beam seen
by the refocusing system in each point of the detector plane is
therefore reduced from $\rband$ to a smaller effective bandwidth
$\band=\rband \Delta X/\Delta X_{\indrm{R}}$.  Using
Eqs.~\eqref{vsp020}-\eqref{vsp030}, it can be presented as
\begin{equation}
\band=
\left|\frac{\dcomm{\an}}{\fcomm{\an}}\right|  \raccept . 
\label{vsp040} 
\end{equation}

Of all the incident photons on the sample,  the spectrometer can
therefore use only those within the effective spectral bandwidth
$\band$, rather than within $\dband$.  In this regard, it is also
important to note that although the effective spectral bandwidth of
the incident photons is reduced to $\band$ because of a limited
angular acceptance $\raccept$ of the dispersing
element D$_{\indrm{\an}}$, the spectral window of imaging is still
intact and equal to the spectral bandwidth $\rband$ of the refocusing
system.  The gain in signal strength of an echo spectrometer compared
to a conventional narrow-band scanning spectrometer with the same
spectral resolution $\Delta\varepsilon$ can be therefore estimated as
\begin{equation}
{\cal G} = \frac{\band}{\Delta\varepsilon}\,\frac{\rband}{\Delta\varepsilon}.
\label{vsp050} 
\end{equation}

Assuming the angular acceptance of D$_{\indrm{\an}}$ is $\raccept
\simeq 250~\mu$rad (see Table~\ref{tab-xes}), we obtain $\Xsize\simeq 50~\mu$m, $\band=10$~meV,  ${\cal G} = 142$ 
for spectrometer \esi , and $\Delta X\simeq 105~\mu$m, $\band=2$~meV,  ${\cal G} = 1600$ 
for spectrometer \esoi .

If a smaller than $\Xsize$ vertical beam size on the sample is
required, it can be always made by installing a beam-defining aperture
in front of the sample. This will reduce proportionally  the signal strength
in the detector but leave intact the performance of the
x-ray echo spectrometer in terms of spectral resolution and the spectral
window of imaging. A better solution is obtained using an angular slit instead
of the aperture, i.e., a Bragg reflecting crystal or a channel-cut
crystal installed after dispersing element  D$_{\indrm{\mo}}$, as
was employed in \cite{SSM13}.

\section{Echo spectrometer tolerances}
\label{sec-tolerances}
Permissible limits of variation of the echo spectrometer parameters
can be calculated from the refocusing condition tolerance given by
Eq.~\eqref{tolerances}.
The latter can be rewritten as
\begin{equation}
|G_{\indrm{\mo}}+G_{\indrm{\an}}/A_{\indrm{\an}}| \band \ll \Delta x_{\ind{1}}
\label{tolerances2} 
\end{equation}
using Eq.~\eqref{refocus} and the relationship $\Delta x_{\ind{2}}=
|A_{\indrm{\an}} | \Delta x_{\ind{1}}$ from Eq.~\eqref{magnification}.
The tolerance intervals can be defined more specifically by setting the requirement 
\begin{equation}
|G_{\indrm{\mo}}+G_{\indrm{\an}}/A_{\indrm{\an}}| \band \lesssim \nu
\Delta x_{\ind{1}}, \hspace{0.5cm} \nu\simeq 0.458,
\label{tolerances27} 
\end{equation}
that limits the blur of the image on the detector and therefore the degradation of
the spectral resolution to 10\%: $\sqrt{1+\nu^2}=1.1$.

In a particular case of the echo spectrometer, which has the optical scheme 
shown in Fig.~\ref{fig002}, the tolerances on the spectrometer
parameters can be calculated by
\begin{equation}
\left| \fcomm{\mo}  \frac{l_{\ind{3}} l_{\ind{1}}}{\dcomm{\mo}^2 l_{\ind{12}}} - \frac{  \fcomm{\an} f_{\ind{1}}\!  }{ \dcomm{\an}\!} \right|  \band \lesssim \nu \Delta x_{\ind{1}} ,
\label{tolerances22} 
\end{equation}
which is obtained combining Eq.~\eqref{tolerances27} and Eqs.~\eqref{defocusing}-\eqref{refocusing}. 
If the dispersing element D$_{\indrm{\mo}}$ is placed from the source
at a large distance $l_{\ind{1}} \gg \dcomm{\mo}^2 l_{\ind{2}}$,  the tolerance equation in
this case simplifies to
\begin{equation}
\left| \fcomm{\mo} l_{\ind{3}}  - \frac{  \fcomm{\an} f_{\ind{1}}\!  }{ \dcomm{\an}\!} \right|  \band \lesssim \nu \Delta x_{\ind{1}} .
\label{tolerances3} 
\end{equation}
As an example, we assume that the spectrometer parameters are
perfectly adjusted, except for the distance $l_{\ind{3}}$ from the
focusing optic to the secondary source (i.e., to the sample). The
tolerance interval $\Delta l_{\ind{3}}$  in this case can be estimated
using Eq.~\eqref{tolerances3} as
\begin{equation}
\left|\Delta l_{\ind{3}} \right| \lesssim \nu \frac{\Delta x_{\ind{1}}}{ |\fcomm{\mo}|  \band  }.
\label{toleranceintervall3} 
\end{equation}
The focal length of  the collimating optic in practice may deviate
from the design value $f_{\ind{1}}$ due to uncertainties in
fabrication. The tolerance interval $\Delta f_{\ind{1}}$ can be
estimated in this case as
\begin{equation}
\left|\Delta  f_{\ind{1}} \right| \lesssim \nu \frac{\Delta x_{\ind{1}} \dcomm{\an}}{ |\fcomm{\an}|  \band  }.
\label{toleranceintervalf1} 
\end{equation}
With the parameters of the 0.1-meV-resolution echo spectrometer
\esoi\ (see Table~\ref{tab-xes}), these tolerance intervals are
estimated to be $|\Delta l_{\ind{3}} | \lesssim 46$~mm and $|\Delta
f_{\ind{1}} | \lesssim 9$~mm.  For the 1-meV-resolution echo
spectrometer \esi , they are $|\Delta l_{\ind{3}} | \ll 91$~mm and
$\left|\Delta f_{\ind{1}} \right| \lesssim 9$~mm.  These requirements
are not extremely demanding.

The variations in $l_{\ind{3}}$ and $f_{\ind{1}}$ can result from the
sample position displacement, provided the sample is very thin, or
from a sample having elongation along the beam and substantial
scattering length of x~rays in the sample, or from uneven sample
shape. Therefore, the above estimated numbers also provide constraints
on the scattering length in the sample and the sample shape and size.

The spectral window of imaging can be technically shifted by varying
the glancing angle of incidence (Bragg angle) of the D~crystal(s) in the
dispersing element of the defocusing system, as discussed in
Sec.~\ref{iwindow}. Such variations, however, simultaneously change
the dispersion rate $\fcomm{\mo}$ of the defocusing system.  How much can
$\fcomm{\mo}$ be changed without substantial violation of the
refocusing condition? From Eq.~\eqref{tolerances3} we find that the
tolerance interval in this case is equal to
\begin{equation}
\left|\Delta \fcomm{\mo} \right| \lesssim \nu \frac{\Delta x_{\ind{1}}}{
  l_{\ind{3}} \band }.
\label{toleranceintervalf5} 
\end{equation}
With the parameters of the 0.1-meV-resolution echo spectrometer
\esoi , we obtain that $|\Delta \fcomm{\mo} | \lesssim
0.64~\mu$rad/meV.  For the 1-meV-resolution \esi\ spectrometer,
$\left|\Delta \fcomm{\mo} \right| \ll 0.18~\mu$rad/meV.  The
permissible shifts of the spectral window of imaging will be discussed in
Sec.~\ref{iwindow} using these tolerance intervals.

If the spectrometer parameters are outside the tolerance intervals
defined by
Eqs.~\eqref{toleranceintervall3}--\eqref{toleranceintervalf5}, the
refocusing condition should be adjusted, as described in the following
section, Sec.~\ref{tuningup}.

\section{Refocusing condition adjustment} 
\label{tuningup}
The optical elements of the echo spectrometer have to be manufactured
with a high accuracy so that the dispersion rates $\fcomm{\mo} ,
\fcomm{\an} $, the asymmetry parameters $\dcomm{\mo} , \dcomm{\an}, $
and the focal distances $f, f_{\ind{1}} $ are within the tolerance
intervals defined by the refocusing condition  Eq.~\eqref{tolerances22}.
This, however, may not always be possible in practice.  To overcome this
problem, the refocusing condition can be exactly matched by adjusting the
distances $l_{\ind{1}}$, $l_{\ind{2}}$, and $l_{\ind{3}}$ in the
defocusing system (see Fig.~\ref{fig002}) leaving all other parameters
of the defocusing and refocusing systems intact. Given that the
source-to-sample distance $l=l_{\ind{1}}+l_{\ind{2}}+l_{\ind{3}}$, as
well as the focal distance
$f=l_{\ind{12}}l_{\ind{3}}/(l_{\ind{12}}+l_{\ind{3}})$, and crystal
parameters are fixed, the distances $l_{\ind{1}}$, $l_{\ind{2}}$, and
$l_{\ind{3}}$ are defined from the above-mentioned constraints, by
solving the equations
\begin{align}
\label{tuning0}
l_{\ind{1}}+l_{\ind{2}}+l_{\ind{3}}& = l\\
\frac{l_{\ind{1}}}{\dcomm{\mo}^2}+l_{\ind{2}}& = \frac{f l_{\ind{3}}}{l_{\ind{3}}-f},
\label{tuning1}
\end{align}
together with the refocusing condition given by  Eq.~\eqref{refocus2}.
The solution of the system of Eqs.~\eqref{refocus2}, \eqref{tuning0}, and  \eqref{tuning1} is 
\begin{equation}
l_{\ind{3}}=\frac{l}{2}-\sqrt{\left(\frac{l}{2}\right)^2-W\left(1-\frac{1}{\dcomm{\mo}^2}\right)-lf},
\label{tuning100}
\end{equation}
\begin{equation}
l_{\ind{1}} = \frac{W}{l_{\ind{3}}-f}, \hspace{1cm} W=f_{\ind{1}}f \frac{\fcomm{\an} \dcomm{\mo}^2 }{\fcomm{\mo}  \dcomm{\an}},
\label{tuning033}
\end{equation}
\begin{equation}
l_{\ind{2}} = \frac{l_{\ind{3}}f-W/\dcomm{\mo}^2}{l_{\ind{3}}-f}. 
\label{tuning030}
\end{equation}
Examples of distances $l_{\ind{3}}$, $l_{\ind{2}}$, and $l_{\ind{1}}$
calculated for slightly varying values of
$\fcomm{\mo}$ and $\dcomm{\mo}$ using Eqs.~\eqref{tuning100},
\eqref{tuning033}, and \eqref{tuning030} are shown in
Tables~\ref{tab10} and \ref{tab11} for the cases of the
1-meV-resolution spectrometer \esi\ and the 0.1-meV-resolution
spectrometer \esoi , respectively.  It is noteworthy that a small
variation in
$\fcomm{\mo}$ and $\dcomm{\mo}$ results in a
small variation of $l_{\ind{3}}$, but in a rather large variation of
$l_{\ind{2}}$.

\begin{table}[t!]
\centering
\begin{tabular}{|lll|lll|l|}
  \hline   \hline 
$\eta_{\indrm{D}}$    & $\fcomm{\mo}$  & $\dcomm{\mo}$  & $l_{\ind{3}}$  & $l_{\ind{2}}$  & $l_{\ind{1}}$ & $A_{\ind{\mo}}$ \\[-5pt]    
 & &  &  &   & & \\[0pt]    
deg   & $\frac{\mu {\mathrm {rad}}}{\mathrm {meV}}$  &  & m  & m & m & \\[5pt]    
\hline  \hline  
83.8 & 2.96 & 1.93 & 1.721  & 0.168 & 33.11 & -0.0984 \\[0pt]    
84.0 & 3.11 & 1.98 & 1.725  & 0.609 & 32.66 & -0.0974 \\[0pt]
84.2 & 3.27 & 2.04 & 1.732  & 1.027 & 32.44 & -0.0967 \\[0pt]        
\hline  \hline  
\end{tabular}
\caption{Distances $l_{\ind{3}}$, $l_{\ind{2}}$, and $l_{\ind{1}}$
  between the optical elements of the defocusing system calculated
  by Eqs.~\eqref{tuning100}-\eqref{tuning030} for slightly different $\eta_{\indrm{D}}$ and
  therefore $\fcomm{\mo}$ and $\dcomm{\mo}$ values with
  $l=l_{\ind{3}}+l_{\ind{2}}+l_{\ind{1}}=35$~m 
  fixed. Other parameters are also fixed and  given in Tables~\ref{tab-1meV} and \ref{tab-xes}. The 1-meV-resolution
  spectrometer \esi\ is considered.
}       
\label{tab10}
\end{table}

\begin{table}[t!]
\centering
\begin{tabular}{|lll|lll|l|}
  \hline   \hline 
$\eta_{\indrm{D}}$    & $\fcomm{\mo}$  & $\dcomm{\mo}$  & $l_{\ind{3}}$  & $l_{\ind{2}}$ & $l_{\ind{1}}$ & $A_{\ind{\mo}}$ \\[-5pt]    
 & &  &  & &  &  \\[0pt]    
deg   & $\frac{\mu {\mathrm {rad}}}{\mathrm {meV}}$  &  & m  & m & m &\\[5pt]    
\hline  \hline  
77.5 & -29.86 & 1.9 & 1.710  & 0.197 & 33.092 & -0.0961\\[0pt]        
78 & -31.65 & 1.96 & 1.716  & 0.730 & 32.553  & -0.0951\\[0pt]
79 & -35.92 & 2.08 & 1.722  & 1.731 & 31.546  & -0.0918\\[0pt]    
\hline  \hline  
\end{tabular}
\caption{Same as Table~\ref{tab10}, however for the 0.1-meV-resolution echo spectrometer \esoi .}
\label{tab11}
\end{table}

There is always a possibility of hitting the limit $l_{\ind{2}}=0$,
which, however, should be avoided in practice.  This suggests the
necessity of an iterative approach in the optical design of the x-ray
echo spectrometers.  In the first step, the initial values of the
parameters entering $W$ in Eq.~\eqref{tuning033} are determined from
the required energy resolution and the refocusing condition, as in
Sec.~\ref{optical-scheme}. In the next step, $W$, $l_{\ind{3}}$, and
$l_{\ind{2}}$ are calculated from
Eqs.~\eqref{tuning100}-\eqref{tuning030}. If $l_{\ind{2}}$ is not
positive, the crystal parameters $\fcomm{\mo}$ and $\dcomm{\mo}$ of
the defocusing dispersing element have to be adjusted to move
$l_{\ind{2}}$ into a comfortable range, e.g., $l_{\ind{2}}= $ 0.25~m to
 1~m. See examples presented in Tables~\ref{tab10} and \ref{tab11}.

\section{Spectral window of imaging and scanning range}
\label{iwindow} 

\begin{figure}[t!]
\includegraphics[width=0.5\textwidth]{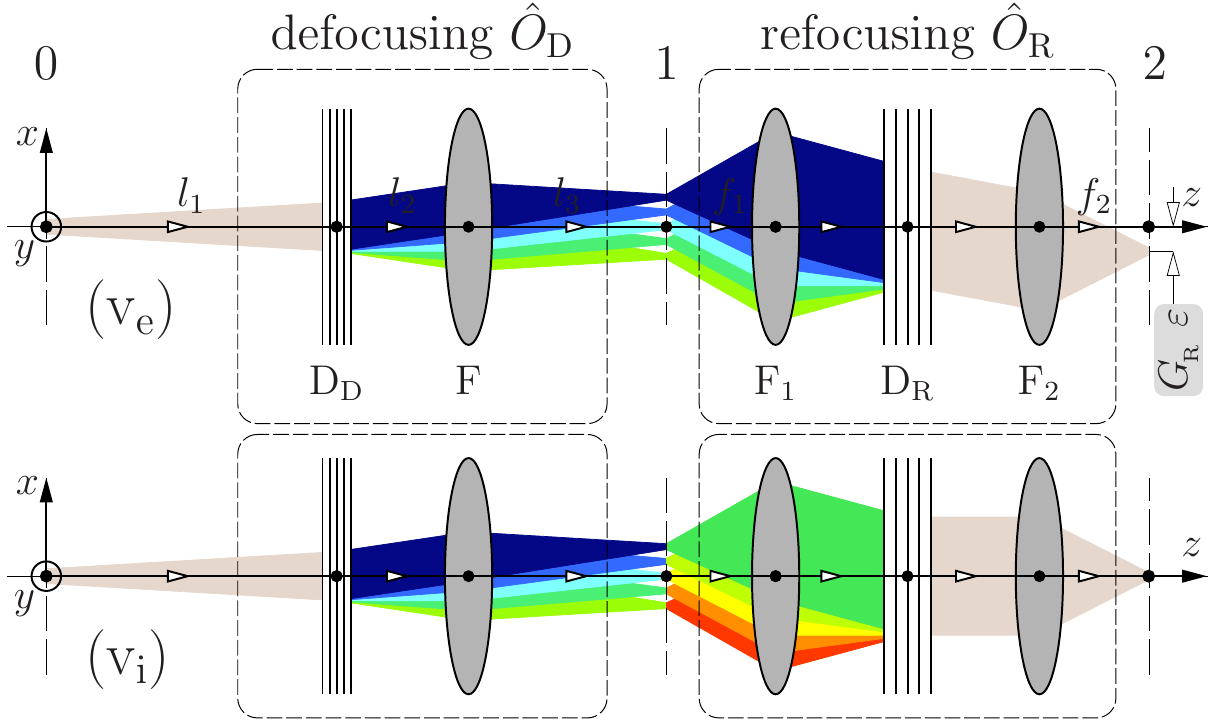}
\caption{Shifting the spectral window of imaging. A change in the crystal
  parameters in the dispersing element D$_{\indrm{\mo}}$ of the
  defocusing system $\hat{O}_{\indrm{\mo}}$ results in a change of the
  spectral composition of x~rays on the sample (compare with
  Fig.~\ref{fig002}). Because the crystal parameters in the dispersing
  element D$_{\indrm{\an}}$ of the refocusing system
  $\hat{O}_{\indrm{\an}}$ are not changed, the elastically scattered
  photons are now refocused on the detector with a spatial shift
  (v$_{\indrm{e}}$). This results in a shift of the spectral imaging
  window. In particular, the inelastically scattered photons can now
  be refocused on the detector into the position which was associated
  before with elastically scattered photons (v$_{\indrm{i}}$).}
\label{fig002b}
\end{figure}

Unlike the conventional scanning-type narrow-band hard x-ray IXS
spectrometers, x-ray echo spectrometers are imaging spectrographs.
The spectral window of imaging, however, is limited and defined by the
bandwidth $\rband$ of the refocusing system. How does one proceed if
IXS spectra have to be imaged with an energy transfer $\varepsilon$
outside the window of imaging?  This can be accomplished by shifting
the window of imaging into the region of interest. The practically
simplest way is to shift the bandwidth $\dband$ of the defocusing
system.  Nothing has to be changed in the refocusing system
$\hat{O}_{\indrm{\an}}$, as illustrated in
Fig.~\ref{fig002b}. Technically, the bandwidth of the defocusing
system can be shifted either by varying the angle of incidence
$\theta$ of the x~rays to the D~crystals of the dispersing elements, or by
varying the crystal temperature and therefore the crystal lattice
parameter.\footnote{The spectral profile of the window of imaging can be
  measured by detecting the elastically scattered signal and scanning
  the bandwidth of the refocusing system in a similar way.}


It is important that the variation of the crystal parameters, e.g.,
the incidence angle $\theta$ of the dispersing element
D$_{\indrm{\mo}}$, does not change the linear dispersion rate in the
defocusing system $\hat{O}_{\indrm{\mo}}$ over the limit $\left|\Delta
\fcomm{\mo} \right|$ determined from
Eq.~\eqref{toleranceintervalf5}, and therefore does not result in a
violation of the refocusing condition. Otherwise, the refocusing
condition has to be readjusted, as discussed in Sec.~\ref{tuningup}.

Let us determine how much of the bandwidth of the defocusing system
can be shifted by varying the glancing angle of incidence $\theta$
without violation of the refocusing condition.  The maximal spectral
shift can be calculated apparently as
\begin{equation}
\delta E_{\indrm{max}} = \pm \left| \frac{{\mathrm d} \fcomm{\mo} }{{\mathrm d}\theta} \frac{{\mathrm d}\theta}{{\mathrm d}E}  \right|^{-1} \left|\Delta \fcomm{\mo} \right|.
\label{ad01037}
\end{equation}

In the simplest case of the dispersing element D$_{\indrm{D}}$
consisting of one asymmetrically cut crystal, which is, e.g., the case
of the 1-meV-resolution spectrometer \esi\ (see Fig.~\ref{fig0012})
$\fcomm{\mo} =\dirate $ given by Eq.~\eqref{ad010}, we can calculate
\begin{equation}
\frac{{\mathrm d}\dirate }{{\mathrm d}\theta} \simeq -\frac{\dirate  }{\theta-\eta}.       
\label{ad01033}
\end{equation}
From Bragg's law,  ${{\mathrm d}\theta}/{{\mathrm d}E}=-\tan\theta /E $. As a result,
\begin{equation}
\delta E_{\indrm{max}} =  \pm E \,
 \frac{\theta-\eta}{\tan\theta}\, \frac{|\Delta \dirate |}{|\dirate|} .
\label{ad01039}
\end{equation}

For the four-crystal CDDW-type dispersing element, the cumulative
dispersion rate $\fcomm{4}$ can be approximated to a good accuracy by
Eq.~\eqref{scum020}. Assuming $\theta_{\ind{3}}$ is close to $\pi/2$
and $\theta_{\ind{3}}-\eta_{\ind{3}}$ is small, the variation of
$\fcomm{4}$ with the glancing angles of incidence $\theta_{\ind{2}}$
and $\theta_{\ind{3}}=\theta_{\ind{2}}$ in Bragg diffraction from the
D~crystals is given by
\begin{equation}
\frac{{\mathrm d}\fcomm{4}}{{\mathrm d}\theta_{\ind{3}}}\simeq -\frac{2\fcomm{4}}{\theta_{\ind{3}}-\eta_{\ind{3}}},
\label{ad0103}
\end{equation}
an expression which is similar to Eq.~\eqref{ad01033}, differing only by a factor of $2$.

For the CDDW optic, the variation of the bandwidth position
with angle ${\mathrm d} E/{\mathrm d}\theta_{\ind{3}} \simeq -E\,
/\tan\theta_{\ind{3}}$ (in the approximation $\tan\theta_{\ind{3}}\gg
\tan\theta_{\ind{1}}$) can be shown. Using this expression together with
Eqs.~\eqref{ad0103} and \eqref{ad01037}, we obtain for the permissible
shift of the spectral window of imaging of the CDDW-type dispersing
element D$_{\indrm{\mo}}$:
\begin{equation}
\delta E_{\indrm{max}} = \pm E\,\frac{\theta_{\ind{3}}-\eta_{\ind{3}}}{2\tan\theta_{\ind{3}}}\, \frac{|\Delta \fcomm{4}|}{|\fcomm{4}|}.
\label{ad0104}
\end{equation}
To a factor of $1/2$, it is equivalent to the one-crystal dispersing
element case given by Eq.~\eqref{ad01039}.

Using Eqs.~\eqref{ad01039} and \eqref{ad0104}, the tolerance interval
values $|\Delta \fcomm{\mo}|$ calculated in Sec.~\ref{sec-tolerances}
for the 1-meV-resolution spectrometer \esi\  and the 0.1-meV-resolution
spectrometer \esoi ,  together with the appropriate values of the
dispersing element parameters from Tables~\ref{tab-1meV} and
\ref{tab-0o1meV}, respectively, we estimate for the permissible shift of
the spectral window of imaging $\delta E_{\indrm{max}} \simeq \pm
0.6$~eV for both spectrometers.  The scanning ranges of the echo-type
spectrometers are relatively large and comparable with those of the
conventional scanning-type IXS spectrometers \cite{Baron16}.


Since $|\delta E_{\indrm{max}}|$ is much larger than the spectral
window of imaging $\rband$, the maximal energy transfer which can be
measured is $E_{\indrm{M}} \simeq |\delta E_{\indrm{max}}|$. It is
very important to note that $|\delta E_{\indrm{max}}|$ and therefore
$E_{\indrm{M}}$ can be substantially increased, if the refocusing
condition adjustment procedure is applied, as described in
Sec.~\ref{tuningup}.

\section{Impact of the secondary source size on the spectral resolution}
\subsection{Vertical secondary source size} 

The vertical {\em monochromatic} secondary source size
is determined by the vertical monochromatic focal spot size $\Delta x_{\ind{1}}$ on
the sample. We assume in the first approximation that they are equal
and do not change with scattering angle $\Phi$ (see Fig.~\ref{fig002}),
provided the collimating optic and subsequent optical components of
the refocusing system are correctly aligned in the scattering plane.
The smallness of $\Delta x_{\ind{1}}$ and more precisely of its
angular size $\Delta x_{\ind{1}}/f_{\ind{1}}$ is critical for
achieving high spectral resolution, given by Eq.~\eqref{resolution2}.

\subsection{Horizontal secondary source size} 
\label{Horizontal-secondary-source-size} 

\begin{figure}
\includegraphics[width=0.5\textwidth]{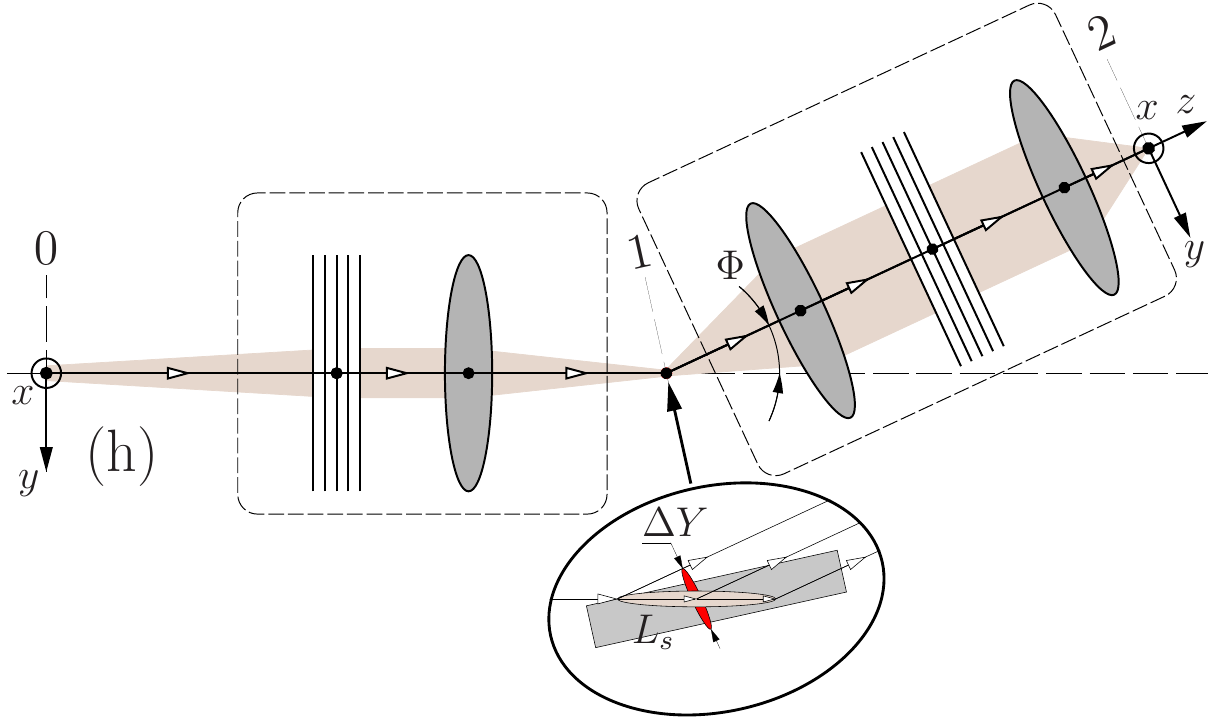}
\caption{Optical scheme of an x-ray echo spectrometer in the
  horizontal scattering plane, the same as in Fig.~\ref{fig002}(h),
  however, showing a close-up of the sample (gray rectangle) and the
  trace of the scattering path in the sample (pink ellipse), which
  details an increase of the secondary horizontal source size $\ysize$
  in the sample with scattering angle $\Phi$. }
\label{fig0015}
\end{figure}

In contrast, the horizontal secondary source size $\ysize $ changes
with the scattering angle $\Phi$ (assuming the horizontal focal spot
size is smaller).  It increases with $\Phi$ as $\ysize \simeq
L_{\indrm{s}}\sin\Phi$ because the projection of the scattering length
$L_{\indrm{s}}$ on the scattering direction increases; see pink
ellipse in Fig.~\ref{fig0015}.  To consider the impact of the
horizontal size on the spectral resolution, we assume for simplicity in
the following that the secondary source is concentrated in the sample
reference plane $1$, as indicated by the red ellipse in
Fig.~\ref{fig0015}, i.e., there is no longitudinal component, and the
secondary source distribution in reference plane $1$ is presented by
coordinates $(x_{\ind{1}}, y_{\ind{1}})$. Such an approximation is well
founded, because the spectral resolution is quite insensitive to the
spread of the secondary source size along the optical axis, as
discussed in Sec.~\ref{sec-tolerances}.

X~rays from secondary source point $(x_{\ind{1}},y_{\ind{1}})$
propagate after the collimating optic $F_{\ind{1}}$ at an angle
$\varphi=y_{\ind{1}}/f_{\ind{1}}$ to the dispersion plane $(x,z)$.
The glancing angle of incidence $\theta_{\ind{1}}$ to the Bragg
reflecting atomic planes of the first crystal of the dispersing
element D$_{\indrm{\an}}$ changes with $\varphi$ to
$\theta_{\ind{1\varphi}}$, where
\begin{equation}
  \sin\theta_{\ind{1\varphi}}=\sin\theta_{\ind{1}} \cos\varphi.
\label{eq00110}
\end{equation}
In the approximation of a small
$\varphi$, the angular difference 
\begin{equation}
\theta_{\ind{1\varphi}}-\theta_{\ind{1}}=-\xi_{\ind{1\varphi}}=\frac{\varphi^2}{2} \tan\theta_{\ind{1}}. 
\label{eq00120}
\end{equation}
After $n$ Bragg reflections (at the exit of the dispersing element
D$_{\indrm{\an}}$ of the refocusing system), the vertical angular
difference $\xi_{\ind{n\varphi}}^{\prime}$ between the direction of
x~rays propagating in the dispersing plane and the direction of x~rays
propagating with an angular deviation $\varphi$ off the plane is
\begin{equation}
\xi_{\ind{n\varphi}}^{\prime} \simeq  \Xi_{\ind{n}} \,\, \xi_{\ind{1\varphi}}.
\label{eq00126}
\end{equation}
The magnitude of $\Xi_{\ind{n}}$ in Eq.~\eqref{eq00126} depends on the
concrete optical design of the dispersing element D$_{\indrm{\an}}$.
In the particular case of the four-crystal ($n=4$) CDDW-type optic in the
($\pi+$,$\pi+$,$\pi-$,$0-$) scattering configuration presented in
Figs.~\ref{fig004v4} and \ref{fig004v3}, $\Xi_{\ind{4}}$ is given by
\begin{equation}
\Xi_{\ind{4}}\, \simeq \, -\frac{b_{\ind{4}} (1- b_{\ind{3}}  b_{\ind{2}})}{\tan\theta_{\ind{1}}\cos\theta_{\ind{2}}}
\label{eq00142}
\end{equation}
as shown in Appendix~\ref{Xi}.  Here $\theta_{\ind{2}}$ is the nominal
glancing angle of incidence to the reflecting atomic planes of the
second crystal, which is assumed to be close to $90^{\circ}$ and equal
to $\theta_{\ind{3}}$; $b_{\ind{2}}, b_{\ind{3}}$, and $b_{\ind{4}}$
are the asymmetry factors of the Bragg reflections from the second,
third, and fourth crystals, respectively.

We assume that the imaging optic F$_{\indrm{2}}$, with a focal
distance $f_{\indrm{2}}$, focuses both vertically and
horizontally. Each point of the secondary source with coordinates
$(x_{\ind{1}},y_{\ind{1}})$ in reference plane $1$ will be imaged
to a point $(x_{\ind{2}},y_{\ind{2}})$ on the detector reference plane $2$, where
\begin{equation}
x_{\ind{2}}=A_{\indrm{\an}} x_{\ind{1}}+\xi_{\ind{n\varphi}}^{\prime}f_{\ind{2}} , \hspace{0.5cm} 
y_{\ind{2}}=-y_{\ind{1}}\frac{f_{\ind{2}}}{f_{\ind{1}}},
\label{eq00132}
\end{equation}
and $A_{\indrm{\an}}=-{\dcomm{\an}f_{\ind{2}}\!}/{f_{\ind{1}}\!}$ is the
magnification factor of the refocusing system in the vertical
dispersing plane; see Eq.~\eqref{refocusing}. We note that in the
horizontal nondispersing plane, the magnification factor is just
$-{f_{\ind{2}}}/{f_{\ind{1}}}$. Using Eqs.~\eqref{eq00120},
\eqref{eq00126},  and \eqref{refocusing}, we obtain from Eq.~\eqref{eq00132}
\begin{equation}
x_{\ind{2}}=\left[x_{\ind{1}}+ U \,\frac{\varphi^2}{2}\right] A_{\indrm{\an}}  , \hspace{0.5cm} 
y_{\ind{2}}=-y_{\ind{1}}\frac{f_{\ind{2}}}{f_{\ind{1}}}
\label{eq00135}
\end{equation}
where 
\begin{equation}
U=\frac{f_{\ind{1}} \Xi_{\ind{n}} \tan\theta_{\ind{1}} }{\dcomm{\an}}. 
\label{eq00136}
\end{equation}
In the particular case of the CDDW-type optic, with $\Xi_{\ind{4}}$ given by \eqref{eq00142}, we obtain
\begin{equation}
U=\frac{f_{\ind{1}} (1- b_{\ind{3}}  b_{\ind{2}}) }{|b_{\ind{1}}b_{\ind{2}}  b_{\ind{3}}| \cos\theta_{\ind{2}}}. 
\label{eq001366}
\end{equation}

\begin{figure}
\includegraphics[width=0.5\textwidth]{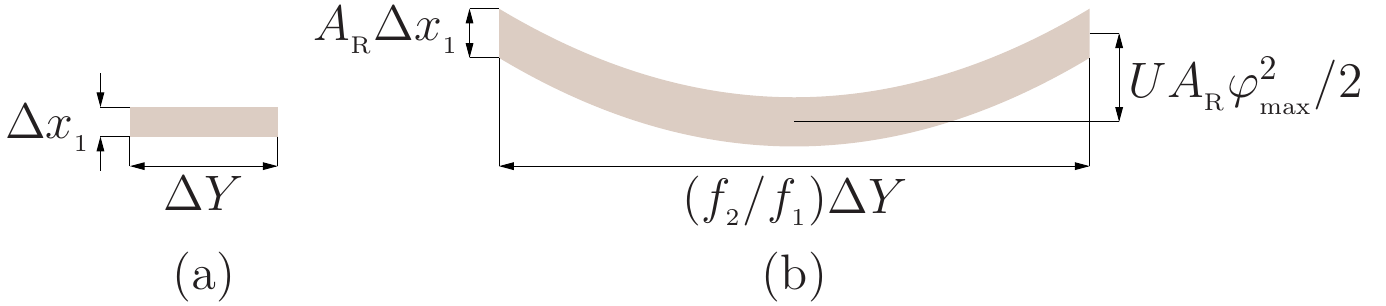}
\caption{The secondary monochromatic source of a rectangular shape in
  reference (sample) plane $1$ (a) being imaged onto reference
  (detector) plane $2$ acquires a curved shape (b) [see
  Eq.~\eqref{eq00135}] with $\varphi_{\indrm{max}}=
  \ysize/2f_{\ind{1}}$.  }
\label{fig0016}
\end{figure}

If the secondary source size has a rectangular shape in reference
(sample) plane $1$ with a height $\Delta x_{\ind{1}}$ and a width
$\ysize$, its image, according to Eq.~\eqref{eq00135}, acquires a
curved shape; see Fig.~\ref{fig0016}. This result is in agreement with
numeric simulations performed in \cite{SC16} for the particular case of
the x-ray echo spectrometer with designed parameters given in
\cite{Shvydko16}.

\subsubsection{Curved image flattening}
\label{rduction-to-flat}

If a 2D-pixel detector is used to record the image, and if $U$ is
known, the curved shape can be reduced numerically to a flat one. $U$
can be determined experimentally from the curvature of the elastic
scattering image. With the curved image reduced to a flat one, the
vertical size reduces to $\Delta x_{\ind{2}}= A_{\indrm{\an}} \Delta
x_{\ind{1}}$, i.e., to a value unaffected by the horizontal source
size. Therefore, if the flattening procedure is applied, the horizontal
source size in the first approximation {\em does not deteriorate the
  spectral resolution}  of the x-ray echo spectrometer.

\subsubsection{Curved image}

In contrast, if a 1D-pixel detector is used, sensitive in the $x$ direction and
integrating in the $y$ direction, the vertical image size $\Delta
x_{\ind{2}}$ increases to
\begin{equation}
\Delta x_{\ind{2}} = |A_{\indrm{\an}}| \sqrt{\xsize ^2\,+  \,\xphisize ^2} 
\label{eq00137}
\end{equation}
\begin{equation}
\xphisize = U \sqrt{\left<\left( \frac{\varphi^2}{2}\right)^2 \right>} = \frac{U \varphi_{\indrm{max}}^2}{2\sqrt{5}} = \frac{U \ysize^2}{8\sqrt{5} f_{\ind{1}}^2}, 
\label{eq00138}
\end{equation}
where $\varphi_{\indrm{max}}= \ysize/2f_{\ind{1}}$. 

The spectral resolution $\Delta \etra$ of the echo spectrometer scales
with the vertical image size $\Delta x_{\ind{2}}$; see
Eq.~\eqref{resolution}. Because of the horizontal spread $\ysize$ of the
secondary source size, $\Delta x_{\ind{2}}$ acquires an additional component
$A_{\indrm{\an}}\xphisize$, resulting in a total vertical source size of
$A_{\indrm{\an}}\sqrt{\xsize ^2 + \xphisize^2 }\simeq A_{\indrm{\an}}\xsize(1+\nu)$, where
$\nu\simeq \xphisize^2/2\xsize^2$. For the spectral resolution not
to  deteriorate by more than $\nu$, we require that $\xphisize
\lesssim \sqrt{2\nu}\xsize$. Combining this expression with
Eq.~\eqref{eq00138}, we obtain for the permissible horizontal secondary source size:
\begin{equation}
\ysize \lesssim v \sqrt{\frac{ 8f_{\ind{1}}^2 \xsize}{U}}, \hspace{0.5cm} v=(10\nu)^{1/4}.
\label{eq00140}
\end{equation}
With $U$ defined in Eq.~\eqref{eq00136}, 
\begin{equation}
\ysize \lesssim v \sqrt{\frac{ 8f_{\ind{1}} \dcomm{\an} \xsize }{\Xi_{\ind{n}} \tan\theta_{\ind{1}} }} 
\label{eq00141}
\end{equation}
in a general case, or with $U$ defined in Eq.~\eqref{eq001366}, 
\begin{equation}
\ysize \lesssim v \sqrt{8f_{\ind{1}} \xsize\, |b_{\ind{1}}b_{\ind{2}}b_{\ind{3}}| \cos\theta_{\ind{2}}/(1-b_{\ind{2}}b_{\ind{3}})}
\label{eq00143}
\end{equation}
for the case of  the CDDW optic.

For our exemplary echo-type IXS spectrometers (see
Tables~\ref{tab-1meV}, \ref{tab-0o1meV}, and \ref{tab-xes}), we
estimate $\ysize \lesssim 185~\mu$m
($\varphi_{\indrm{max}}=460~\mu$rad; $U=46$~m) for 1-meV-resolution
spectrometer \esi\ and $\ysize \lesssim 120~\mu$m
($\varphi_{\indrm{max}}=150~\mu$rad; $U=450$~m) for 0.1-meV-resolution
spectrometer \esoi , assuming a 10\% limit ($\nu=0.1$ and $v=1$) of
the spectral resolution degradation.


\section{Focusing optics}  

Focusing optics is another group of key elements of the x-ray echo
spectrometer.  A distinctive feature of the echo-type spectrometers is
the propagation of different spectral components at different
sometimes large angles to the optical axis; see Fig.~\ref{fig002}. The
angular deviation from the optical axis can be as large as
$\pm\rdtheta/2 \simeq \pm 150~\mu$rad (see Tables~\ref{tab-1meV},
\ref{tab-0o1meV}, and \ref{tab-xes}) and may result in coma
aberrations, and therefore degradation of the spectral resolution. It
is essential that the focusing (F), collimating (F$_{\ind{1}}$), and
imaging (F$_{\ind{2}}$) optical elements of the  echo
spectrometer, are capable of producing sharp images both with on-axis
as well as off-axis illumination, i.e., they should be truly
aberration-free imaging optical elements.

X-ray compound-refractive parabolic lenses (CRL) are genuine imaging devices
\cite{LST99} and are appropriate for x-ray echo spectrometers.
However, because of the large photo absorption and therefore small,
typically less than 1~mm effective apertures, their application is
limited perhaps to the focusing element F of the defocusing system.

Grazing incidence curved mirrors have higher efficiency and therefore
seem to be a preferred choice, especially for collimating
(F$_{\ind{1}}$) and imaging (F$_{\ind{2}}$) optical elements. In the
first approximation, they may have 2D paraboloidal shapes.  Such
mirrors are becoming available now \cite{YKM16}. Alternatively, more
traditional systems composed of two 1D parabolic mirrors mounted at
$90^{\circ}$ to each other can be used as well. Kirkpatrick-Baez (KB)
mirrors \cite{KB48} are arranged in-line one after the other, while
Montel mirrors \cite{Montel} are mounted side by side. Montel optics
are especially attractive when the focal distance is comparable with
the mirror length, which is the case for collimating elements
F$_{\ind{1}}$.

Grazing incidence paraboloidal (parabolic) mirrors can focus x-ray
beams properly, but only those propagating parallel to the parabola
axis.  A parallel beam with a lateral size $\beamsize$ propagating at
an angular deviation $\xi$ from the axis is focused to a spot enlarged
due to coma to a size of \footnote{Xianbo Shi (APS) private
  communication.}
\begin{equation}
w\,=\,\beamsize\,{|\xi|}/{\gamma},
\label{shinbo}
\end{equation}
and shifted by $x=\xi f$ from the optical axis.  Here $\gamma$ is 
nominal grazing incidence angle and $f$ is the mirror's focal length. To
prevent spectral resolution degradation, it is essential that coma
$w\ll \Delta x_{\ind{i}}$, i.e., much smaller than the perfect
monochromatic image size $\Delta x_{\ind{i}}$ on the sample ($i=1$)
and on the detector ($i=2$).

Typically, $\gamma\simeq 3$~mrad for grazing incidence mirrors
designed for $\simeq 9$~keV x~rays. For our exemplary echo-type IXS
spectrometers (see Table~\ref{tab-xes}) $|\xi| \lesssim 150~\mu$rad and
$\beamsize \gtrsim 1$~mm. Thus, coma can enlarge the focal spots to $w\simeq
50~\mu$m and more, i.e., to sizes which are more or much more than
$\Delta x_{\ind{i}}$.  Therefore, grazing incidence parabolic mirrors as they
are cannot be used as focusing optics of x-ray echo spectrometers.

\subsection{Mitigating coma}
\label{mitigatingcoma}

Coma $w$ of a paraboloidal mirror, Eq.~\eqref{shinbo}, can be
mitigated if the incidence angle $\gamma$ can be enlarged and the
incident x~ray's beam size $\beamsize$ can be reduced. The angle
$\gamma$ can be enlarged by an order of magnitude and $w$ reduced by
the same amount, if one employs graded multilayer mirrors.  Indeed,
state-of-the-art commercially available  high-reflectivity
($\simeq 70$\%) Mo/B$_{\ind{4}}$C graded-multilayer mirrors designed
for 9-keV photons may feature $\gamma\simeq 30$~mrad
\cite{Platonov17}. Additionally, in the particular case
of mirror F$_{\ind{2}}$, the problem can be mitigated further by
increasing the focal distance $f_{\ind{2}}$, which is yet a free
parameter, and therefore by increasing $\Delta x_{\ind{2}}$; see
Eqs.~\eqref{magnification} and \eqref{refocusing}. Let us verify that
this may work in the particular cases of the exemplary spectrometers.

{\bf Imaging mirror F$_{\ind{2}}$.} The imaging element F$_{\ind{2}}$
in the refocusing dispersing system $\hat{O}_{\indrm{\an}}$ focuses
x~rays onto the detector. Because the vertical beam size after the
dispersing element D$_{\indrm{\an}}$ can be as large as
$\beamsize=\beamsize_{\indrm{\an}}/\dcomm{\an}=3.5$~mm (see
Table~\ref{tab-xes}) the imaging element F$_{\ind{2}}$ has to have as
large vertical geometrical aperture $A_{\indrm{2g}}$. Paraboloidal
mirrors with graded multilayer coatings and a large incidence angle
$\gamma\simeq 30$~mrad may feature sufficient geometrical aperture,
relatively small length $\simeq A_{\indrm{2g}}/\gamma$, and relatively
small coma $w < \Delta x_{\ind{2}}$.

Indeed, we estimate $w=14~\mu$m, for the case of the 1-meV-resolution
spectrometer \esi , assuming $|\xi|=\pm \rdtheta/2= \pm 0.12$~mrad and
the above-mentioned values of $\gamma$ and $B$.  For the
0.1-meV-resolution \esoi\ spectrometer with
$\beamsize=\beamsize_{\indrm{\an}}/\dcomm{\an}=0.5$~mm, we obtain
$w=2.3~\mu$m.  The estimated $w$ values are a factor of two to three
smaller than the appropriate monochromatic image sizes $\Delta
x_{\ind{2}}$ given in Table~\ref{tab-xes}. Therefore, coma may
degrade the spectral resolution in these cases by less that 10\%.

{\bf Collimating mirror F$_{\ind{1}}$.}  The collimating element
F$_{\ind{1}}$ in the refocusing dispersing system
$\hat{O}_{\indrm{\an}}$ collects photons in a large solid angle
$\Omega_{\indrm{h}}\times\Omega_{\indrm{v}}$, with
$\Omega_{\indrm{h}}\simeq \Omega_{\indrm{v}}\simeq 1-10$~mrad
(depending on the required momentum transfer resolution $\Delta
Q\simeq K \Omega_{\indrm{h}}$), and makes parallel x-ray beams of each
spectral component (assuming point secondary source). Laterally graded
multilayer Montel mirrors proved to be useful exactly in this role
\cite{MSL13,SSS14,SCC14}.

The impact of the coma aberration in collimating mirror
F$_{\indrm{1}}$ on the spectral resolution of the echo spectrometer
$\Delta \etra = {|A_{\indrm{\an}}|\Delta x_{\ind{1}}
}/{|G_{\indrm{\an}}|} $ [Eq.~\eqref{resolution}] can be estimated by
propagating parallel monochromatic beams in the opposite direction
and calculating the effective increase of the ideal monochromatic
secondary source size $\Delta x_{\ind{1}}$ due to coma $w$, given by
Eq.~\eqref{shinbo}.

In particular, for the \esi\ spectrometer with $f_{\ind{1}}=0.2$~m,
$\xi \simeq \Xsize/2f_{\ind{1}} = 0.125$~mrad, assuming a lateral
size of the monochromatic collimated beam $\beamsize =\beamsize_{\indrm{\an}} =1$~mm 
and a grazing
incidence angle $\gamma=30$~mrad, we obtain $w=4.1~\mu$m. Such coma
increases by 30\% the effective monochromatic secondary source size
from $\Delta x_{\ind{1}}=5~\mu$m to an effective value of
$\sqrt{\Delta x_{\ind{1}}^2+w^2}$, resulting also in a 30\% degradation of
the spectral resolution.

For the \esoi\ spectrometer, with $f_{\ind{1}}=0.4$~m, $\xi \simeq
\Xsize/2f_{\ind{1}} = 0.125$~mrad, and $\beamsize=\beamsize_{\indrm{\an}}=0.5$~mm
(0.06~nm$^{-1}$ resolution), we obtain $w=2.1~\mu$m. Such broadening
(coma) results in an 8\% degradation of the spectral resolution.

The above examples demonstrate that increasing the incidence angle
$\gamma$ may substantially reduce coma.

\subsection{Aberration-free optics}

Single-reflection mirrors like grazing incidence paraboloidal mirrors
suffer from coma, preventing true imaging, as already discussed in the
previous section.  Aberration-free imaging of an extended source or imaging over some
extended field, involving off-axis mirror illumination, requires at
least two reflections from two reflecting surfaces which exactly
obey the Abbe sine condition \cite{BornWolf,Aschenbach09,Harvey10}.

Wolter optic, composed of two grazing incidence mirrors, is able to
create an x-ray imaging system with a relatively wide field of view
\cite{Wolter52}. Wolter systems typically consist of a paraboloidal
primary mirror and a hyperboloidal or ellipsoidal secondary
mirror. Wolter optics still may suffer from coma aberrations.  To
eliminate coma completely, small corrections are required to the
mirror profile from their nominal second-order shape \cite{Wolter52b}.

Combined KB-Wolter systems were proposed \cite{KKI96} and realized
\cite{MEK15} for applications at synchrotron and x-ray free-electron
laser sources.  More advanced systems are under consideration
\cite{YMS17} for full-field spectroscopy applications.  Such KB-Wolter
systems can be used as the aberration-free focusing element F of the
defocusing system of the x-ray echo spectrometers.

\begin{figure}
\includegraphics[width=0.5\textwidth]{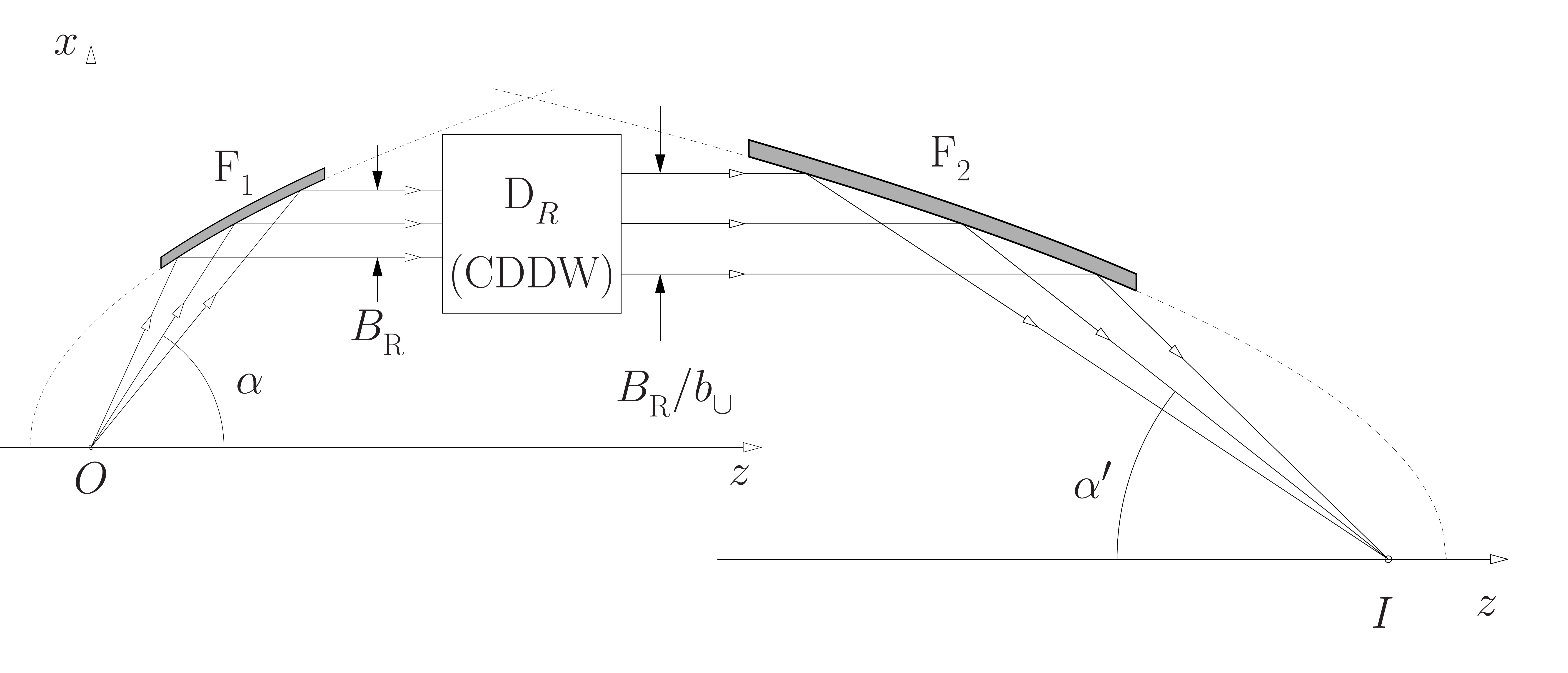}
\caption{Optical scheme of the refocusing system of the
  x-ray echo spectrometer, see Fig.~\ref{fig002}, here shown 
  comprising parabolic collimating mirror F$_{\ind{1}}$, parabolic
  imaging mirror F$_{\ind{2}}$, and dispersing element
  D$_{\indrm{\an}}$ (CDDW-type four-crystal system;
  Figs.~\ref{fig004v4} and \ref{fig004v3}) in between. The Abbe sine
  condition is fulfilled exactly ($\sin\alpha/\sin\alpha^{\prime}=1$)
  in the case of  one-to-one imaging, which takes place if the focal
  distances of the mirrors are related as
  $f_{\ind{2}}=f_{\ind{1}}/\dcomm{\an}$. Here
  $f_{\ind{1}}=O$F$_{\ind{1}}$,
  $f_{\ind{2}}=I$F$_{\ind{2}}$. }
\label{fig0018}
\end{figure}

The refocusing system already comprises two mirrors for collimation
and imaging the secondary source on the detector. The question arises
as to whether such a two-mirror system could be an aberration-free
Wolter-type one.  Such a possibility was already considered by Howells
with regard to soft x-ray plane grating monochromators
\cite{Howells80}, spectroscopic instruments with an optical scheme
very similar to the discussed here scheme of the refocusing system of
the echo spectrometer.  Unlike the original proposal of Wolter,
Howells suggested a double-mirror system in a parabola-parabola
configuration.  Such a configuration has the great advantage of
producing parallel x~rays between the two reflections at the
parabolas, see Figs.~\ref{fig0018} and \ref{fig0018a}, which is
perfect for the proper performance of a plane diffraction grating
inserted between the collimating and focusing mirrors in the
monochromator scheme. The parallel rays between the collimating and
focusing mirrors in the parabola-parabola configuration are also
perfect for the CDDW-type ``plane diffraction gratings'' considered in
the present paper for the x-ray echo spectrometers.

In Appendix~\ref{abbe}, we show that the Abbe sine condition is
satisfied exactly for a system with two identical parabolas producing
one-to-one imaging. In a more general case of a system with two
arbitrary parabolas, the sine condition may be satisfied to a very good
accuracy, in particular, in cases of interest for x-ray echo
spectrometer applications.

It should be noted, however, that because the CDDW dispersing crystal
element changes the cross section of the x-ray beam from
$B_{\indrm{\an}}$ to $B_{\indrm{\an}}/\dcomm{\an}$, the 1:1 imaging
with magnification factor $|A_{\indrm{\an}}|=1$ takes place if the
focal distance of the imaging mirror
$f_{\ind{2}}=f_{\ind{1}}/\dcomm{\an}$, see Eq.~\eqref{refocusing},
differs from the focal distance of the collimating mirror
$f_{\ind{1}}$ by a factor of $1/\dcomm{\an}$. In other words, the
identical parabolas, ensuring perfect imaging obeying Abbe sine
condition under these conditions, cannot be actually identical. The
focal distance and the size of the second mirror should be scaled by
the same factor $1/\dcomm{\an}$ as the beam cross section; see
Fig.~\ref{fig0018}.
  
\section{Pixel detectors and spectral resolution of echo spectrometers}

The monochromatic image size $\Delta x_{\ind{2}}$ on the pixel
detector determines the required spatial resolution of the detector,
which should be $\Delta x_{\indrm{D}}\ll \Delta x_{\ind{2}}$, to not
deteriorate the echo spectrometer spectral resolution $\Delta \etra =
\Delta x_{\ind{2}}/|G_{\indrm{\an}}|$; see Eq.~\eqref{resolution}.  If
the spatial resolution of the detector cannot be neglected, i.e.,
$\Delta x_{\indrm{D}} \gtrsim \Delta x_{\ind{2}}$, the spectral
resolution degrades to
\begin{equation}
\Delta\etra = {\sqrt{(\Delta x_{\ind{2}})^2+(\Delta x_{\indrm{D}})^2}}/{|G_{\indrm{\an}}|}.
\label{resolution3}
\end{equation}
In the particular case of the echo spectrometer with the
optical scheme presented in Sec.~\ref{optical-scheme} and
in Fig.~\ref{fig002}, the spectral resolution given by
Eq.~\eqref{resolution2} is transformed using Eq.~\eqref{resolution3} to
\begin{equation}
\Delta \etra = \frac{|\dcomm{\an}|}{|\fcomm{\an}|}\,\frac{\Delta x_{\ind{1}}}{f_{\ind{1}}}\,\sqrt{1+\left(\frac{1}{|\dcomm{\an}|} \frac{\Delta x_{\indrm{D}}\,f_{\ind{1}}}{\Delta x_{\ind{1}}\, f_{\ind{2}}}\right)^2}.
\label{resolution4}
\end{equation}
Equation~\eqref{resolution4} indicates that a large $f_{\ind{2}}$ is
beneficial for diminishing the negative impact of the limited spatial
resolution.  If a less than 10\% spectral resolution degradation is
permissible, we estimate from Eq.~\eqref{resolution4} for the required
detector spatial resolution: $\Delta x_{\indrm{D}}=15~\mu$m and
$\Delta x_{\indrm{D}}=3~\mu$m for our exemplary echo-type IXS
spectrometers \esi\ and \esoi , respectively (see parameters in
Table~\ref{tab-xes}). Appropriate for this application, x-ray
photon-counting detectors with $\Delta x_{\indrm{D}}=2~\mu$m are state
of the art \cite{SBD12,DBC14}.


\begin{table*}[t!]
\begin{tabular}{|l|l|l||l|} \hline
~~~~~~~~~~~~~~~~~~~~~~~~~~~~~~~~~~~~~~~~~~~~~~~~~~Spectrometer:  & \esi  & \esoi  & HERIX        \\ 
Parameter:  &       &           &   \\ \hline\hline
Photon energy $E$ [keV] & 9.137 &  9.137 & 23.74  \\
Photon momentum $K$ [nm$^{-1}$] & 46.3 & 46.3  & 120.3 \\ \hline
Spectral resolution $\Delta \varepsilon$  [meV] & 1  & 0.1 & 1.5   \\
Momentum transfer resolution $\Delta Q$  [nm$^{-1}$] & 0.46 & 0.05   & 1.2  \\ 
Angular acceptance $\Omega_{\indrm{v}}\times\Omega_{\indrm{h}}$  [mrad$^2$] & 10$\times$10   & 1.1$\times$1.1 & 10$\times$10   \\ 
Effective bandwidth $\band$  [meV] & 10 &  2 & 1   \\
Spectral window of imaging $\rband$  [meV] & 14.2  & 8.0 & 1   \\
Max. energy transfer $E_{\indrm{M}}$  [eV] \footnote{Can be substantially increased if the  adjustment procedure of the
refocusing condition is applied; see Sec.~\ref{tuningup}.}   & 0.6 & 0.6 & 10   \\
Max. scattering angle $\Phi_{\indrm{M}}$ & $ 154^{\circ}$  &  $ 154^{\circ}$ &  $ 35 ^{\circ}$ \\
Max. momentum transfer $Q_{\indrm{M}}$ [nm$^{-1}$] & 90 &  90  & 70  \\ 
Analyzer arm length [m]&  3  &   3.5  & 9  \\
Incident photon polarization & $\pi$ &  $\pi$ &  $\sigma$  \\
Effective vertical beam size on the sample $\Xsize$ [$\mu$m] & 50 & 100 & 20 \\ 
Permissible horizontal secondary sources size $\ysize$ [$\mu$m] \footnote{Can be substantially increased by curved image flattening procedure; see Sec.~\ref{rduction-to-flat}.}  &  185 &   120 \color{black}  & 1000 \\ \hline
Cumulative $\fcomm{\mo}$ of D$_{\indrm{\mo}}$ [$\mu$rad/meV] & -3.12 &  -31.7 & -  \\
Cumulative $\dcomm{\an}$ of D$_{\indrm{\an}}$ & 0.65   & 0.27 & -  \\
Cumulative $\fcomm{\an}/\dcomm{\an}$ of D$_{\indrm{\an}}$ [$\mu$rad/meV] & -25 &  -125 & -  \\
Angular acceptance $\raccept$ of D$_{\indrm{\an}}$  [$\mu$rad] & 246  & 262 & -  \\
Angular divergence $\rdtheta=\rband|\fcomm{\an}|$ after D$_{\indrm{\an}}$  [$\mu$rad] & 234  & 272 & -  \\
Vertical beam size $\beamsize_{\indrm{\an}}$ on D$_{\indrm{\an}}$ [mm] & 2  & 0.43 &-  \\
Vertical beam size $\beamsize_{\indrm{\an}}/\dcomm{\an}$ after D$_{\indrm{\an}}$ [mm] & 3.5  & 1.6  &-  \\
Focusing mirror focal length $f$ [m]  & 1.446 \footnote{A CRL composed of 17 double-convex Be
  lenses with $R=200~\mu$m.} & 1.446   & 1.5  \\
Collimating mirror  focal length  $f_{\ind{1}}$ [m]& 0.2  & 0.4 & -  \\
Imaging mirror  focal length  $f_{\ind{2}}$ [m]& 2 & 2 & -  \\
Smallest image size $\Delta x_{\ind{2}}$ on the pixel detector [$\mu$m]& 32 & 6.8  & -  \\ \hline
Spectral flux $F$  \footnote{As predicted for the  standard undulator of the upgraded Advanced Photon Source  \cite{APSU}. } [ph/meV/s] $\times 10^{10}$ & 30 & 30 & 4.2 \\ \hline
Relative signal strength $S/S^{\indrm{HERIX}}$ & 1014$\times\zeta$ & 1.3 $\times\zeta$  & 1   \\ \hline  
\end{tabular}
\caption{Operation parameters and performance characteristics of the
  considered exemplary echo spectrometers \esi\ and
  \esoi\ compared with the parameters of the state-of-the-art
  narrow-band scanning IXS spectrometer HERIX \cite{TAS11,SSD11}.  The
  signal strength is $S \propto \band \times \rband \times
  \Omega_{\indrm{v}}\times\Omega_{\indrm{h}} \times F\times
  L_{\indrm{s}}$, where $L_{\indrm{s}}$ is the scattering length. The
  relative signal strength values in the bottom row have to be
  corrected for each particular sample using
  $\zeta=L_{\indrm{s}}/L_{\indrm{s}}^{\indrm{HERIX}}$.  The scattering
  length $L_{\indrm{s}}$ is given either by the absorption length
  $L_{\indrm{a}}$ or by the sample thickness, if it is smaller.
  Typically $L_{\indrm{a}}/L_{\indrm{a}}^{\indrm{HERIX}}\simeq 1/2 -
  1/30$.  A {\em monochromatic} focal spot size of $\Delta
  x_{\ind{1}}=5~\mu$m on the sample is assumed in all cases. }
\label{tab-xes}
\end{table*}

\section{Discussion and Conclusions}

Hard x-ray echo spectroscopy, a space-domain counterpart of neutron
spin echo, was recently introduced \cite{Shvydko16} to overcome
limitations in the spectral resolution and weak signals of the traditional
narrow-band scanning IXS probes. X-ray echo spectroscopy relies on
imaging IXS spectra, and does not require x-ray monochromatization. Due
to this, the echo-type IXS spectrometers are broadband and have a
potential to simultaneously provide dramatically increased signal
strength, reduced measurement times, and higher resolution compared to
the traditional narrow-band scanning-type IXS spectrometers.  The main
components of the x-ray echo spectrometer are defocusing and
refocusing systems, composed of dispersing and focusing elements.

The theory of the x-ray echo spectrometers presented in
\cite{Shvydko16} is developed here further with a focus on questions
of practical importance, which could facilitate optical design and
assessment of the feasibility and performance of echo spectrometers.
Among others, the following questions are addressed: spectral
resolution, refocusing condition, echo spectrometers tolerances,
refocusing condition adjustment, effective beam size on the sample,
spectral window of imaging and scanning range, impact of secondary
source size on the spectral resolution, angular dispersive optics,
focusing and collimating optics, and detector's spatial resolution.

The analytical ray-transfer matrix (ray-tracing) approach is used to
calculate spectral resolution, refocusing condition, echo
spectrometer tolerances, etc. Spectral bandwidth and efficiency of
the dispersing elements are calculated using dynamical diffraction
theory of x-ray Bragg diffraction in crystals.

The developed theory provides recommendations on the optical design of
the x-ray echo spectrometer and on the design procedure. In
particular, the equations defining the spectral resolution and the
refocusing condition can be used for the initial estimation of the
dispersion rates required for the dispersing elements, which in turn
determine the possible optical design of the dispersing
elements. Four-crystal CDDW-type arrangements of asymmetrically cut
crystals are proposed as large-dispersion-rate dispersing elements.
The optical parameters of the x-ray echo spectrometers can be further
specified more precisely by refining the refocusing condition. The
refocusing condition is also essential for the calculation of the echo
spectrometer tolerances. If the dispersion rate of a dispersing
element, or the focal length of a focusing element deviates from its
design value, the refocusing condition can be tuned by adjusting the
distances between the optical elements of the defocusing system. This
procedure is very useful, in particular, for extending the spectral
scanning range from a fraction of an eV to a few eV. Another important
recommendation of the theory is to apply the numerical procedure of
flattening the curved image on the detector, and thus to eliminate the
detrimental influence of the horizontal secondary source on the
spectral resolution.

Examples of optical designs and characteristics of echo spectrometers
with 1-meV and 0.1-meV resolutions are discussed in the paper and
supported by the theory. The theory is used to calculate the operation
and performance characteristics of the exemplary x-ray echo
spectrometers, which are summarized in Table~\ref{tab-xes}. These are
compared with what is possible with the state-of-the-art narrow-band
scanning-type IXS spectrometers \cite{Baron16}, in particular with
HERIX, a 1.5-meV-resolution IXS spectrometer at the Advanced Photon
Source (APS) \cite{TAS11,SSD11}.  The signal of the 1-meV-resolution
echo-type spectrometer \esi\ is enhanced by three orders of magnitude
compared to HERIX, provided the scattering length in the sample is the
same; see Table~\ref{tab-xes} for more details. The momentum
resolution is three times better for the same solid angle of
collection of scattered photons.  The signal strength of the
0.1-meV-resolution echo-type spectrometer \esoi\ is comparable to that
of the 1-meV-resolution HERIX spectrometer.  Importantly, not only the
spectral resolution is improved by an order of magnitude; the momentum
transfer resolution of \esoi\ is also improved compared to HERIX by a
factor of 25 (from 1.2~nm$^{-1}$ to 0.05~nm$^{-1}$).  That means that
IXS experiments with an order of magnitude improved spectral and
momentum transfer resolutions are becoming feasible at
storage-ring-based x-ray sources by applying the x-ray echo
spectroscopy approach.

The point is that an even higher spectral resolution $\Delta \etra
\lesssim 0.02$~meV and momentum transfer resolution can be achieved
with x-ray echo spectrometers by increasing the dispersion rates
$\fcomm{}$ in the dispersing elements. This, however, will result in
their narrower transmission bandwidths $\Delta E_{\ind{\cup}}$.
Still, an approximately constant ratio $\Delta E_{\ind{\cup}}/\Delta
\etra$ will hold. Alternatively, the spectral resolution can be
improved by increasing the focal length $f_{\ind{1}}$ of the
collimating optic F$_{\ind{1}}$, or by reducing the secondary source
size $\xsize$ (by improving focusing on the sample); see
Eq.~\eqref{resolution2}. The signal strength will drop.  However,
high-repetition-rate self-seeded x-ray free-electron lasers will
provide in the future orders of magnitude more spectral flux than what
is possible at storage ring sources \cite{CGk16}, and therefore will
make feasible  experiments with an extremely high spectral resolution
$\Delta \etra \lesssim 0.01$~meV.

It is essential that the focusing (F), collimating (F$_{\ind{1}}$),
and imaging (F$_{\ind{2}}$) optical elements of the x-ray echo
spectrometer are capable of producing sharp images with both on-axis
and off-axis illumination; i.e., they should be truly aberration-free
imaging optical elements. The spectral resolution and spectral line
shape will largely depend on the quality of the focusing, collimating,
and imaging optical elements.

The magnitude of the image $\Delta x_{\ind{2}}$ on the pixel detector
defines the required spatial resolution, which is in the
15-$\mu$m to 3-$\mu$m range, depending on the spectrometer. Detectors with
such spatial resolution are state of the art \cite{SBD12,DBC14}.

X-ray echo spectrometers require a combination and precise coupling of
the CDDW dispersing elements and focusing optics as major optical
components. Such coupling and proper functioning of each individually
intricate component, have been experimentally demonstrated recently
\cite{SSM13,SSS14}.  Implementation of x-ray echo spectrometers is,
therefore, realistic.

\section{Acknowledgments}
Lahsen Assoufid, Patricia Fernandez, Jonathan Lang, and Stephan
Streiffer (Advanced Photon Source) are acknowledged for supporting the
project.  Xianbo Shi (Advanced Photon Source) is acknowledged for
valuable discussions of aberrations in x-ray parabolic mirrors.  Work
at Argonne National Laboratory was supported by the US Department of
Energy, Office of Science, Office of Basic Energy Sciences, under
contract DE-AC02- 06CH11357.

\appendix

\begin{table*}
\centering
\caption{Ray-transfer matrices $\{ABG,CDF,001\}$ of optical systems
  used in the paper. The matrices are shown starting with basic ones
  in rows 1--3. Matrices of combined systems are given in rows
  4--6. The ray-transfer matrices of the defocusing
  $\hat{O}_{\indrm{D}}$ and refocusing $\hat{O}_{\indrm{R}}$ systems
  of x-ray echo spectrometers are presented in rows 7 and 8. Definition
  of the glancing angle of incidence $\theta$ to the reflecting
  crystal atomic planes, the asymmetry angle $\eta$, and the
  deflection sign $\sgn$ in Bragg diffraction from a crystal, used for
  the Bragg reflection ray-transfer matrix in row 3, are given in
  Fig.~\ref{fig005}. See Ref.~\cite{Shvydko15} for more details.  }
\begin{tabular}{|l|l|l|l|}
  \hline 
& & & \\
  Optical system & Matrix notation & Ray-transfer matrix  & Definitions and remarks\\[-5pt]    
& & & \\
  \hline  \hline
& & & \\[-3.8mm]  
\parbox[c]{0.25\textwidth}{Free space \cite{KL66,Siegman}\\  \includegraphics[width=0.25\textwidth]{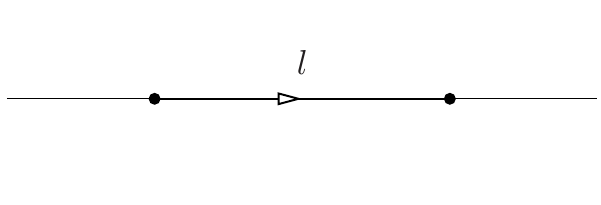}\\[-8mm] \hspace*{-40mm} (1)\\[3mm] }       &  $\hat{\fspace}(l)$ & $\left( \begin{array}{ccc} 1 & l  & 0 \\ 0 & 1 & 0 \\ 0 & 0 & 1  \end{array} \right)$      & \parbox[c]{0.21\textwidth}{$l$ -- distance}     \\[-0.5mm] 
\hline 
& & & \\[-3.8mm]  
\parbox[c]{0.25\textwidth}{Thin lens \cite{KL66,Siegman}\\ \includegraphics[width=0.25\textwidth]{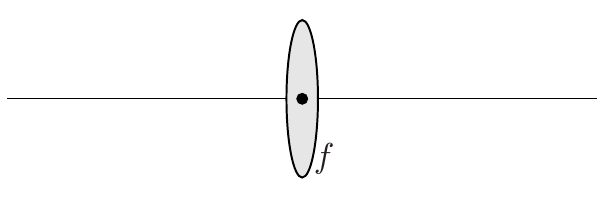}\\[-8mm] \hspace*{-40mm} (2)\\[3mm] }       &  $\hat{\thinlens}(f)$ & $\left( \begin{array}{ccc} 1 & 0 & 0\\ -\frac{1}{f} & 1 & 0  \\ 0 & 0 & 1 \end{array} \right)$      &  \parbox[c]{0.21\textwidth}{$f$ -- focal length}      \\ [-0.5mm]  
\hline 
& & & \\[-4.2mm]  
\parbox[c]{0.25\textwidth}{Bragg reflection from a crystal\\ \cite{MK80-1,MK80-2}\\ \includegraphics[width=0.25\textwidth]{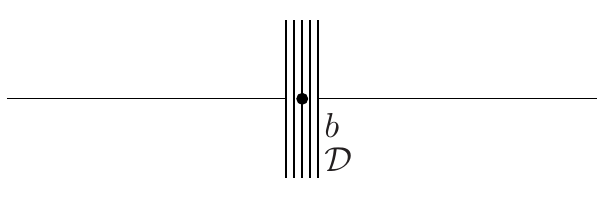} \\[-8mm] \hspace*{-40mm} (3)\\[3mm] }  &  $\hat{\crystal}(b,\sgn \dirate)$ & $\left( \begin{array}{ccc} {1}/{b} & 0 & 0  \\0  & b & \sgn\dirate \\ 0 & 0 & 1 \end{array} \right)$  &  \parbox[c]{0.21\textwidth}{$b=-\frac{\sin(\theta+\eta)}{\sin(\theta-\eta)}$\\ asymmetry factor;\\ $\dirate = -(1/E)(1+b)\tan\theta $\\ angular dispersion rate.
}     \\[+0.5mm] 
\hline 
& & & \\[-3.8mm]  
\parbox[c]{0.25\textwidth}{Successive Bragg reflections \cite{Shvydko15}\\ $\hat{\crystal}(\!b_{\ind{n}},\sgn_{\ind{n}}\dirate_{\ind{n}}\!)\dotsb \hat{\crystal}(\!b_{\ind{1}},\sgn_{\ind{1}}\dirate_{\ind{1}}\!)$ \\  \includegraphics[width=0.25\textwidth]{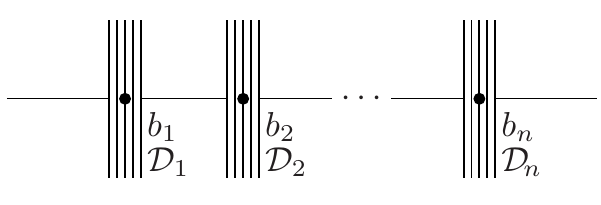} \\[-8mm] \hspace*{-40mm} (4)\\[3mm] }  &  $\hat{\crystal}(\dcomm{n},\fcomm{n})$ & $\left( \begin{array}{ccc} \acomm{n} & 0 & 0  \\ 0  & \dcomm{n}  & \fcomm{n} \\ 0 & 0 & 1 \end{array} \right)$  &  \parbox[c]{0.21\textwidth}{$\dcomm{n}=b_{\ind{1}}b_{\ind{2}}b_{\ind{3}} \dotsc b_{\ind{n}}$ \\ $\fcomm{n}=b_{\ind{n}}\fcomm{n-1} + \sgn_{\ind{n}}\dirate_{\ind{n}}$\\ $\sgn_i=\pm 1$, $i=1,2,...,n$ }     \\[-0.5mm] 
\hline 
& & & \\[-3.8mm]  
\parbox[c]{0.25\textwidth}{Successive Bragg reflections with space between crystals \cite{Shvydko15}\\ $\hat{\crystal}(\!b_{\ind{n}},\sgn_{\ind{n}}\dirate_{\ind{n}}\!)\dotsb \hat{\fspace}(\!l_{\ind{12}}\!)\hat{\crystal}(\!b_{\ind{1}},\sgn_{\ind{1}}\dirate_{\ind{1}}\!)$ \\  \includegraphics[width=0.25\textwidth]{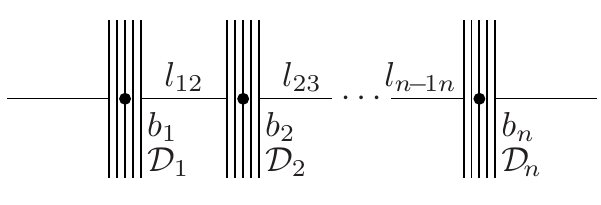}\\[-8mm] \hspace*{-40mm} (5)\\[3mm] }  &  $\hat{\crystalsp}(\dcomm{n},\fcomm{n},l)$ & $\left( \begin{array}{ccc} \acomm{n} & \bcomm{n} & \gcomm{n}  \\ 0  & \dcomm{n}  & \fcomm{n} \\ 0 & 0 & 1 \end{array} \right)$  &  \parbox[c]{0.21\textwidth}{$\bcomm{n}\!=\!\frac{\bcomm{n-1}+\dcomm{n-1}l_{\ind{n-1 n}}}{b_{\ind{n}}}$\\[1mm]
$\gcomm{n}\!=\!
\frac{\gcomm{n-1}+\fcomm{n-1}l_{\ind{n-1 n}}}{b_{\ind{n}}}$\\[1mm]  $\bcomm{1}\!=\!0, \hspace{0.5cm} \gcomm{1}\!=\!0 $}     \\[-0.5mm] 
\hline 
& & & \\[-3.8mm]
\parbox[c]{0.25\textwidth}{Focusing system\\ $\hat{\fspace}(l_{\ind{2}})\hat{\thinlens}(f)\hat{\fspace}(l_{\ind{1}})$ \\ \includegraphics[width=0.25\textwidth]{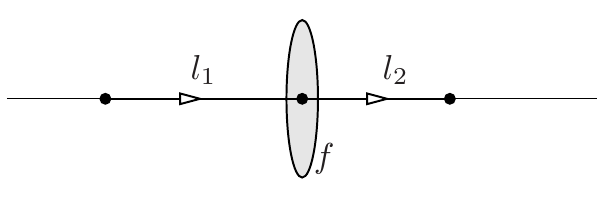} \\[-8mm] \hspace*{-40mm} (6)\\[3mm] }  &  $\hat{\focus}(l_{\ind{2}},f,l_{\ind{1}})$ & $\left( \begin{array}{ccc} 1-\frac{l_{\ind{2}}}{f} & B_{\indrm{F}} & 0  \\ -\frac{1}{f}  & 1-\frac{l_{\ind{1}}}{f} & 0\\ 0 & 0 & 1  \end{array} \right)$  &  \parbox[c]{0.21\textwidth}{$B_{\indrm{F}}=l_{\ind{1}}l_{\ind{2}}\left(\frac{1}{l_{\ind{1}}}+\frac{1}{l_{\ind{2}}}-\frac{1}{f}\right)$ }     \\[-0.5mm] 
\hline 
\hline 
& & & \\[-3.8mm]
\parbox[c]{0.25\textwidth}{Defocusing system $\hat{O}_{\indrm{D}}$ \cite{Shvydko15}\\ $\hat{\focus}(l_{\ind{3}},f,l_{\ind{2}}) \hat{\crystal}(\dcomm{n}\!,\fcomm{n}\!)\hat{\fspace}(l_{\ind{1}})$ \\ \includegraphics[width=0.25\textwidth]{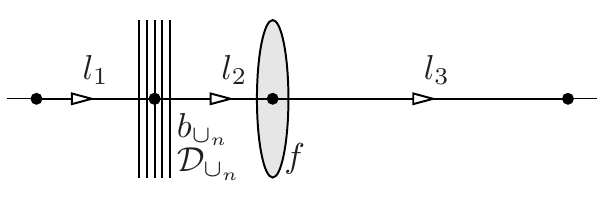} \\[-8mm] \hspace*{-40mm} (7)\\[3mm] }  &  $\hat{O}_{\indrm{D}}$ & $\left(\!\! \begin{array}{ccc} \frac{1}{\dcomm{n}\!}\!\left(\!1\!-\!\frac{l_{\ind{3}}}{f}\!\right) & 0  &  X \fcomm{n}  \\ 
-\frac{1}{f\dcomm{n}}  & \dcomm{n}\!\!\left(\!1\!-\!\frac{l_{12}}{f}\!\right) & \left(\!1\!-\!\frac{l_{\ind{2}}}{f}\!\right)\!\fcomm{n} \\ 
0 & 0 & 1  \end{array}\!\! \right)
 $  &  \parbox[c]{0.21\textwidth}{$\frac{1}{l_{\ind{12}}}\!+\!\frac{1}{l_{\ind{3}}}\!=\!\frac{1}{f}$ \\[1mm]
$l_{\ind{12}} =l_{\ind{1}} /\dcomm{n}^2 + l_{\ind{2}}  $ \\[1mm] $X\!=\!l_{\ind{3}} l_{\ind{1}}\!/(\!\dcomm{n}^2\! l_{\ind{12}}\!) $  }     \\[-0.5mm] 
\hline 
& & & \\[-3.8mm]
\parbox[c]{0.25\textwidth}{Refocusing system $\hat{O}_{\indrm{R}}$ \cite{Shvydko15}\\ $\hat{\focus}(\!f_{\ind{2}}\!,\!f_{\ind{2}}\!,\!l_{\ind{2}}\!) \hat{\crystal}(\!\dcomm{n}\!,\!\fcomm{n}\!) \hat{\focus}(\!l_{\ind{1}}\!,\!f_{\ind{1}}\!,\!f_{\ind{1}}\!) $\\ \includegraphics[width=0.25\textwidth]{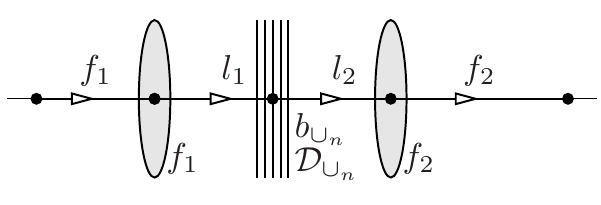} \\[-8mm] \hspace*{-40mm} (8)\\[3mm] }   &  $\hat{O}_{\indrm{R}}$ & $ 
\left(\!\! \begin{array}{ccc} -\frac{\dcomm{n}f_{\ind{2}}\!}{f_{\ind{1}}\!}  & 0 &  f_{\ind{2}} \fcomm{n}  \\ \frac{(\!l_{\ind{1}}\!-\!f_{\ind{1}}\!)\!+\!(\!l_{\ind{2}}\!-\!f_{\ind{2}}\!)\dcomm{n}^2\!}{\!\dcomm{n}f_{\ind{1}}f_{\ind{2}}\!}  & -\frac{f_{\ind{1}}}{\dcomm{n}\!f_{\ind{2}}} & \left(\!1\!-\!\frac{l_{\ind{2}}}{f_{\ind{2}}}\!\right)\!\fcomm{n}\! \\ 0 & 0 & 1 \end{array}\!\! \right)
 $  &  \parbox[c]{0.21\textwidth}{ }     \\ 
  \hline  
\end{tabular}
\label{tab2}
\end{table*}

\section{Ray-transfer matrices} 
\label{ray-transfer-matrices}

Ray-transfer matrices $\{A0G,CDF,001\}$ of the defocusing
$\hat{O}_{\indrm{D}}$ and refocusing $\hat{O}_{\indrm{R}}$ systems of
the x-ray echo spectrometers used in the paper are given in the last
two rows of Table~\ref{tab2}. They are equivalent to the ray-transfer
matrices of x-ray focusing monochromators and spectrographs derived in
Ref.~\cite{Shvydko15}. The matrices of the multielement systems
$\hat{O}_{\indrm{D}}$ and $\hat{O}_{\indrm{R}}$ are obtained by
successive multiplication of the matrices of the constituent optical
elements, which are given in the upper rows of Table~\ref{tab2}.

In the first three rows, 1--3, matrices are shown for the basic
optical elements, such as propagation in free space
$\hat{\fspace}(l)$, thin lens or focusing mirror $\hat{\thinlens}(f)$,
and Bragg reflection from a crystal $\hat{\crystal}(b,\sgn \dirate)$.
Scattering geometries in Bragg diffraction from crystals are defined
in Fig.~\ref{fig005}.  In the following rows of Table~\ref{tab2},
ray-transfer matrices are shown for arrangements composed of several
basic optical elements, such as successive multiple Bragg reflections
from crystals $\hat{\crystal}(\dcomm{n},\fcomm{n})$ and
$\hat{\crystalsp}(\dcomm{n},\fcomm{n},l)$, rows 4--5; and a focusing
system $\hat{\focus}(l_{\ind{2}},f,l_{\ind{1}})$, row 6.

The matrices of the defocusing $\hat{O}_{\indrm{D}}$ and refocusing
$\hat{O}_{\indrm{R}}$ systems presented in Table~\ref{tab2}, rows 7
and 8, respectively, are calculated using the multicrystal matrix
$\hat{\crystal}(\dcomm{n},\fcomm{n})$ from row 4, assuming zero free
space between crystals in successive Bragg reflections.
Generalization to a more realistic case of nonzero distances between
the crystals requires the application of matrix
$\hat{\crystalsp}(\dcomm{n},\fcomm{n},l)$ from row 5.

We refer to Ref.~\cite{Shvydko15} for details on the derivation of
these matrices. Here, we provide only the final results, notations,
and definitions.

\section{Derivation of $\Xi_{\ind{n}}$}
\label{Xi}
In Sec.~\ref{Horizontal-secondary-source-size}, we consider a linear
secondary source in reference plane 1 extended in the horizontal
scattering plane along the $y$~axis (perpendicular to the optical axis
$z$); see Fig.~\ref{fig0015}. Each point of the linear secondary
source radiates x~rays in $4\pi$, but the collimating element
F$_{\ind{1}}$ captures them in a large solid angle and makes them
propagate in parallel towards the dispersing element D$_{\ind{\an}}$.
The rays propagate parallel to the plane $(y,z)$ but at an angle
$\varphi$ to the dispersion plane of the first crystal, which is
parallel to plane $(x,z)$ in Fig.~\ref{fig0015}, and at an angle
$\theta_{\ind{1\varphi}}$ to the diffracting atomic planes of the
first crystal; see Eq.~\eqref{eq00120}. We will consider $n$
successive Bragg reflections from $n$ crystals, and will calculate the
vertical angular difference $\xi_{\ind{n\varphi}}^{\prime}$ after the
$n$th reflection between the direction of x~rays propagating in the
dispersion plane ($\varphi=0$) and the direction of x~rays propagating
with an angular deviation $\varphi$ off the plane. In particular, we
will show that $\xi_{\ind{n\varphi}}^{\prime} \simeq \Xi_{\ind{n}}
\,\, \xi_{\ind{1\varphi}}$ and derive the constant $\Xi_{\ind{n}}$;
see Eq.~\eqref{eq00126}.

For each crystal, we use here a local reference system $\{x_{\ind{\!
    n}}^{\prime},y_{\ind{\!  n}}^{\prime},z_{\ind{\! n}}^{\prime}\}$,
as defined in \cite{Shvydko-SB} (Fig.~2.4).  We assume that the
dispersion planes $(x_{\ind{\!  n}}^{\prime},z_{\ind{\!
    n}}^{\prime})$ of all crystals are parallel to each other (as well
as all $y_{\ind{\!  n}}^{\prime}$~axes).  Wave vectors of photons
incident on and diffracted from the $n$th crystal in this reference
system can be presented by
\begin{equation}
\vc{K}_{\ind{\! n\varphi}}=K\left(\!\! \begin{array}{l} 
\cos\theta_{\ind{\! n\varphi}}\cos\phi_{\ind{\! n\varphi}} \\ 
\cos\theta_{\ind{\! n\varphi}}\sin\phi_{\ind{\! n\varphi}} \\ 
- \sin\theta_{\ind{\! n\varphi}}  \end{array} \!\! \right)\!\!, \hspace{0.05cm}
\vc{K}_{\ind{\! n\varphi}}^{\prime} =  K  \left(\!\! \begin{array}{l} 
\cos\theta_{\ind{\! n\varphi}}^{\prime}\cos\phi_{\ind{\! n\varphi}}^{\prime} \\ 
\cos\theta_{\ind{\! n\varphi}}^{\prime}\sin\phi_{\ind{\! n\varphi}}^{\prime} \\  
\sin\theta_{\ind{\! n\varphi}}^{\prime}  \end{array} \!\! \right)\!\!,
\label{xi00600}
\end{equation}
respectively. Here, $\theta_{\ind{\! n\varphi}}$ is the glancing angle
of incidence and $\theta_{\ind{\! n\varphi}}^{\prime}$ the glancing
angle of reflection measured relative to the diffracting atomic planes
parallel to $(x_{\ind{\! n}}^{\prime},y_{\ind{\! n}}^{\prime})$, while
$\phi_{\ind{\! n\varphi}}$ is the azimuthal angle of incidence and
$\phi_{\ind{\! n\varphi}}^{\prime}$ is the azimuthal angle of
reflection measured as a deviation from the  dispersion plane.

The angular deviation $\varphi$ relates to the
azimuthal angle $\phi_{\ind{\! 1}}$ by
\begin{equation} 
\phi_{\ind{\! 1\varphi}}=\frac{\varphi}{\cos\theta_{\ind{\! 1}}}.
\label{xi00300}
\end{equation}
We assume that $\varphi$, $\phi^{\prime}_{\ind{\! n\varphi}}$, and
$\phi_{\ind{\! n\varphi}}$ are small for all crystals.  It can be shown that under
these conditions to a good accuracy 
\begin{equation} 
\phi^{\prime}_{\ind{\! n\varphi}}\simeq \phi_{\ind{\! n\varphi}}.
\label{xi00400}
\end{equation}

Following the rule that the counterclockwise sense of angular
variations of the ray slope $\xi$ to the optical axis is positive (see
Fig.~3 of \cite{Shvydko15} for more details) we define
\begin{equation} 
\theta_{\ind{\! n\varphi}}=\theta_{\ind{\! n}} - \sgn_{\ind{\! n}} \xi_{\ind{ n\varphi}}, \hspace{0.5cm} \theta_{\ind{\! n\varphi}}^{\prime}=\theta_{\ind{\! n}}^{\prime} + \sgn_{\ind{\! n}}  \xi_{\ind{ n\varphi}}^{\prime}.  
\label{xi00200}
\end{equation}
Here $\theta_{\ind{\! n}}$ and $\theta_{\ind{\! n}}^{\prime}$ are the
nominal ``Bragg angles'' of incidence and reflection, respectively,
of x~rays with a particular photon energy  propagating in the
dispersion planes at $\varphi=\phi=0$.  The angular variations $\xi_{\ind{
    n\varphi}}^{\prime}$ and $\xi_{\ind{ n\varphi}}$ are related to
each other by
\begin{equation} 
\xi_{\ind{ n\varphi}}^{\prime} = b_{\ind{\! n}} \xi_{\ind{ n\varphi}},  
\label{xi00250}
\end{equation}
as follows from the Bragg reflection ray-transfer matrix
$\hat{\crystal}(b,\sgn \dirate)$ (see Table~\ref{tab2}) assuming that
a small deviation $\phi$ of x~rays from   the dispersion plane does not violate it. 

For all crystals to be in Bragg reflection,  each  successive crystal $n$ has to be rotated by an angle
\begin{equation}
\alpha_{\ind{\! n}}\, = \, \left\{ \begin{array}{ll}  
\theta_{\ind{\! n-1}}^{\prime}+\theta_{\ind{\! n}} & ~~~\mbox{in\,}(++)\,\mbox{or\,}(--)\,\mbox{geometry} \\ 
\theta_{\ind{\! n-1}}^{\prime}-\theta_{\ind{\! n}}+\pi & ~~~\mbox{in\,}(+-)\,\mbox{or\,}(-+)\,\mbox{geometry} \\ 
\end{array}\right.
\label{xi00700}
\end{equation}
about the $y_{\ind{\! n}}^{\prime}$ crystal axis, which is parallel to
the $y_{\ind{\! n-1}}^{\prime}$ axis of the previous $(n-1)$th
crystal.  The rotation matrix in this case is
\begin{equation}
\hat{R}(\alpha_{\ind{\! n}})  =  \left(\!\! \begin{array}{ccc} \cos\alpha_{\ind{\! n}}  & 0 &  \sin\alpha_{\ind{\! n}}  \\ 0   & 1  & 0 \\  -\sin\alpha_{\ind{\! n}} & 0 &  \cos\alpha_{\ind{\! n}} \end{array}\!\! \right).
\label{xi00800}
\end{equation}

The momentum of a photon reflected from the $n$th crystal and
incident on the $(n+1)$th crystal can be presented in the reference
system $\{x_{\ind{\! n}}^{\prime},y_{\ind{\!  n}}^{\prime},z_{\ind{\!
    n}}^{\prime}\}$ of the $n$th crystal as $\vc{K}_{\ind{\!
    n,\varphi}}^{\prime}$ and as $\vc{K}_{\ind{\! n+1,\varphi}}$ in
the reference system $\{x_{\ind{\! n+1}}^{\prime},y_{\ind{\!
    n+1}}^{\prime},z_{\ind{\! n+1}}^{\prime}\}$ of the $(n+1)$th
crystal, and related to each other by
\begin{equation} 
\vc{K}_{\ind{\! n+1,\varphi}} =  \hat{R}(\alpha_{\ind{\! n+1}}) \,\,\vc{K}_{\ind{\! n\varphi}}^{\prime}.
\label{xi00705}
\end{equation}
Using  Eq.~\eqref{xi00600} for $\vc{K}_{\ind{\! n\varphi}}^{\prime}$,
and $\vc{K}_{\ind{\! n+1,\varphi}}$  and equalizing vector  components 
in Eq.~\eqref{xi00705}, we have
\begin{widetext}
\begin{align}
\cos\theta_{\ind{\! n+1,\varphi}}\cos\phi_{\ind{\! n+1\varphi}} & = \cos\theta_{\ind{\! n\varphi}}^{\prime}\cos\phi_{\ind{\! n\varphi}} \cos\alpha_{\ind{\! n+1}}   + \sin\theta_{\ind{\! n\varphi}}^{\prime} \sin\alpha_{\ind{\! n+1}} \label{xi01001}, \\
\cos\theta_{\ind{\! n+1,\varphi}}\sin\phi_{\ind{\! n+1,\varphi}}  & = \cos\theta_{\ind{\! n\varphi}}^{\prime}\sin\phi_{\ind{\! n\varphi}}   \label{xi01002}, \\
- \sin\theta_{\ind{\! n+1,\varphi}} & =
-\cos\theta_{\ind{\! n\varphi}}^{\prime}\cos\phi_{\ind{\! n\varphi}} \sin\alpha_{\ind{\! n+1}}  + \sin\theta_{\ind{\! n\varphi}}^{\prime} \cos\alpha_{\ind{\! n+1}}.
\label{xi01000}
\end{align}
\end{widetext}

Taking $\theta_{\ind{\! n+1,\varphi}}=\theta_{\ind{\! n+1}} -
\sgn_{\ind{\! n+1}} \xi_{\ind{ n+1,\varphi}}$ from Eq.~\eqref{xi00200}
and the fact that $|\xi_{\ind{ n+1,\varphi}}|\ll 1$, we can present the left-hand side
of Eq.~\eqref{xi01000} as
\begin{multline}
- \sin\theta_{\ind{\! n+1,\varphi}} \simeq
-\sin\theta_{\ind{\! n+1}} + \sgn_{\ind{\! n+1}} \xi_{\ind{ n+1,\varphi}}\cos\theta_{\ind{\! n+1}} . 
\label{xi01100}
\end{multline}
Using $\theta_{\ind{\! n\varphi}}^{\prime}=\theta_{\ind{\!
    n}}^{\prime} + \sgn_{\ind{\! n}} b_{\ind{\! n}} \xi_{\ind{
    n\varphi}}$ from Eqs.~\eqref{xi00200} and \eqref{xi00250}, the
approximation $\cos\phi_{\ind{\! n\varphi}}\simeq  (1-\phi_{\ind{\! n\varphi}}^2/2$),
Eq.~\eqref{xi00700}, and
omitting terms $\propto \phi^2\xi$, we can present the right-hand side
of Eq.~\eqref{xi01000} as
\begin{multline}
- \cos\theta_{\ind{\! n\varphi}}^{\prime} 
\cos\phi_{\ind{\! n\varphi}} \sin\alpha_{\ind{\! n+1}}\cos\phi_{\ind{\! n+1}}   +  \sin\theta_{\ind{\! n\varphi}}^{\prime}  \cos\alpha_{\ind{\! n+1}} \\
\simeq -\sin\theta_{\ind{\! n+1}} 
+ \sgn_{\ind{\! n}} b_{\ind{\!n}} \xi_{\ind{ n\varphi}} \cos(\theta_{\ind{\!  n}}^{\prime}-\alpha_{\ind{\! n+1}}) +
\frac{\phi_{\ind{\! n\varphi}}^2}{2}\cos\theta_{\ind{\!  n}}^{\prime}\sin\alpha_{\ind{\! n+1}}.
\label{xi01200}
\end{multline}
From Eq.~\eqref{xi00700} we have
\begin{equation}
\cos(\alpha_{\ind{\! n+1}}-\theta_{\ind{\! n}}^{\prime})\, = \, S_{\ind{n+1,n}} \cos\theta_{\ind{\! n+1}}
\label{xi01660}
\end{equation}
where $S_{\ind{n+1,n}}=+ 1$ in the $(++)$ or $S_{\ind{n+1,n}}=- 1$ in the $(+-)$ scattering geometry. 
Finally, using Eq.~\eqref{xi01660} and inserting Eqs.~\eqref{xi01100}--\eqref{xi01660} into Eq.~\eqref{xi01000}, we obtain a recursive relationship for $\xi_{\ind{ n\varphi}}$:
\begin{equation}
\sgn_{\ind{\! n+1}}\xi_{\ind{ n+1,\varphi}}
= S_{\ind{n+1,n}} b_{\ind{\!n}} \sgn_{\ind{\! n}} \xi_{\ind{ n\varphi}}   +
\frac{\phi_{\ind{\! n\varphi}}^2}{2}\frac{\cos\theta_{\ind{\!  n}}^{\prime}}{\cos\theta_{\ind{\! n+1}}}\sin\alpha_{\ind{\! n+1}}.
\label{xi01300}
\end{equation}
A recursive relationship for $\phi_{\ind{\! n\varphi}}$ is derived from
Eq.~\eqref{xi01002}:
\begin{equation}
\phi_{\ind{\! n+1,\varphi}}  \simeq \frac{\cos\theta_{\ind{\! n}}^{\prime}}{\cos\theta_{\ind{\! n+1}}}\,\,\,\phi_{\ind{\! n\varphi}}.
\label{xi01360}
\end{equation}
Now, $\Xi_{\ind{ n}}$ [see Eq.~\eqref{eq00126}] can be calculated
using the above recursive relationships together with
Eq.~\eqref{xi00250}.
 
Here, as an example, we will calculate $\Xi_{\ind{ 4}}$ for the
particular case of the four-crystal ($n=4$) CDDW-type optic in the
($\pi+$,$\pi+$,$\pi-$,$0-$) scattering configuration presented in
Figs.~\ref{fig004v4} and \ref{fig004v3}.

Taking $S_{\ind{2,1}}=+1$,  $\sgn_{\ind{\! 1}}=+1$, 
$\theta_{\ind{\!  1}}^{\prime}\simeq \theta_{\ind{\!  1}}$,
$\sin\alpha_{\ind{\!  2}}=\sin(\theta_{\ind{\! 2}}+\theta_{\ind{\!
    1}}^{\prime})\simeq \cos\theta_{\ind{\! 1}}$ (we assume that
$\theta_{\ind{\! 2}}$ is close to $\pi/2$ as is the case of the
D~crystals $n=2,3$), and ${\phi_{\ind{\! 1\varphi}}}^2/{2}=-{\xi_{\ind{\!
      1\varphi}}}/{(\sin\theta_{\ind{\! 1}}\cos\theta_{\ind{\! 1}})}$
derived from Eqs.~\eqref{eq00120} and \eqref{xi00300}, we have from
Eq.~\eqref{xi01300} that
\begin{equation}
\sgn_{\ind{\! 2}}\xi_{\ind{ 2\varphi}}
= b_{\ind{\!1}} \xi_{\ind{ 1\varphi}}   
-\frac{\xi_{\ind{\! 1\varphi}}}{\tan\theta_{\ind{\! 1}}\cos\theta_{\ind{\! 2}}}.
\label{xi01430}
\end{equation}
Because $|b_{\ind{\!1}}|\ll 1$, $\tan\theta_{\ind{\!  1}}\ll 1$, and $\cos\theta_{\ind{\!  2}}\ll 1$, Eq.~\eqref{xi01430} approximates to
\begin{equation}
\sgn_{\ind{\! 2}}\xi_{\ind{ 2\varphi}} \simeq -\frac{\xi_{\ind{\! 1\varphi}}}{\tan\theta_{\ind{\! 1}}\cos\theta_{\ind{\! 2}}}.
\label{xi01500}
\end{equation}
Further, taking $S_{\ind{3,2}}=-1$, $\theta_{\ind{\!
    2}}^{\prime}\simeq\theta_{\ind{\! 3}}$, and $\sin\alpha_{\ind{\!
    3}}=\sin(\theta_{\ind{\! 2}}^{\prime}-\theta_{\ind{\! 3}}+\pi)
\simeq 0$, we have from Eqs.~\eqref{xi01300} and \eqref{xi01500} that
\begin{equation}
\sgn_{\ind{\! 3}}\xi_{\ind{ 3\varphi}}
= - b_{\ind{\!2}}\sgn_{\ind{\! 2}}  \xi_{\ind{ 2\varphi}}
\simeq \frac{b_{\ind{\!2}} \xi_{\ind{\! 1\varphi}}}{\tan\theta_{\ind{\! 1}}\cos\theta_{\ind{\! 2}}}.
\label{xi01600}
\end{equation}
Similarly, taking $S_{\ind{4,3}}=+1$, $\cos\theta_{\ind{\!  3}}^{\prime}\ll 1$,  and $\sin\alpha_{\ind{\! 4}}=\sin(\theta_{\ind{\! 3}}^{\prime}+\theta_{\ind{\! 4}}) \simeq \sin(\pi/2+\theta_{\ind{\! 4}}) \simeq \cos\theta_{\ind{\! 4}}$, we obtain from Eqs.~\eqref{xi01300} and \eqref{xi01600}:
\begin{multline}
\sgn_{\ind{\! 4}}\xi_{\ind{ 4\varphi}}
=  b_{\ind{\!3}} \sgn_{\ind{\! 3}} \xi_{\ind{ 3\varphi}}   
+\frac{\phi_{\ind{\! 3\varphi}}^2}{2}\frac{\cos\theta_{\ind{\!  3}}^{\prime}}{\cos\theta_{\ind{\! 4}}}\sin\alpha_{\ind{\! 4}}\\
\simeq -\frac{(1-b_{\ind{\!3}}b_{\ind{\!2}}) \xi_{\ind{\! 1\varphi}}}{\tan\theta_{\ind{\! 1}}\cos\theta_{\ind{\! 2}}}.
\label{xi01770}
\end{multline}
Finally, from Eqs.~\eqref{xi01770} and \eqref{xi00250}, we have for the angular spread
$\xi_{\ind{ 4\varphi}}^{\prime}=\theta_{\ind{\! 4\varphi}}^{\prime}-\theta_{\ind{\! 4}}^{\prime}$ 
after the 4th crystal 
\begin{equation}
\xi_{\ind{ 4\varphi}}^{\prime}=
\sgn_{\ind{\! 4}} b_{\ind{\! 4}} \xi_{\ind{ 4\varphi}}=
\Xi_{\ind{\! 4}} \xi_{\ind{\! 1\varphi}}, 
\label{xi01800}
\end{equation}
where
\begin{equation}
\Xi_{\ind{\! 4}} \simeq - \frac{b_{\ind{\! 4}} (1-b_{\ind{\!3}}b_{\ind{\!2}})}{\tan\theta_{\ind{\! 1}}\cos\theta_{\ind{\! 2}}}.
\label{xi01900}
\end{equation}

\section{Abbe sine condition for Wolter-type parabola-parabola optic}
\label{abbe}

\begin{figure}[t!]
\includegraphics[width=0.5\textwidth]{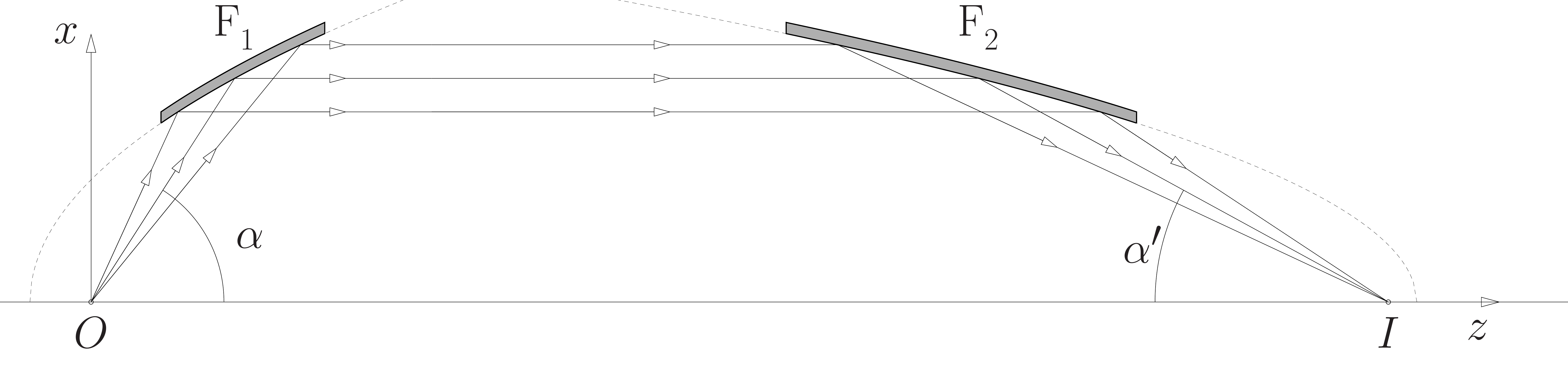}
\caption{Wolter-type double-reflection imaging optic comprising two
  parabolic mirrors F$_{\ind{1}}$ and $F_{\ind{2}}$.}
\label{fig0018a}
\end{figure}

Howells proposed using a Wolter-type double-reflection system for
designing plane grating spectrometers with a good coma-free off-axis
imaging satisfying the Abbe sine condition \cite{Howells80}.  Unlike the
original proposal of Wolter, Howells  suggested a double-mirror
parabola-parabola configuration.  Such a mirror combination has the great
advantage of producing parallel x~rays between the two reflections at
the parabolas, see Figs.~\ref{fig0018} and Fig.~\ref{fig0018a}, which is perfect for the
proper performance of a plane diffraction grating inserted between
the collimating and focusing mirrors in the spectrometer. The parallel
rays between the collimating and focusing mirrors in the
parabola-parabola configuration is also perfect for the CDDW-type
flat-crystal dispersing elements considered in the present paper for
the x-ray echo spectrometers.

Here we show that the Abbe sine condition
\begin{equation}
\sin\alpha/\sin\alpha^{\prime}\,=\,{\mathrm {const}} 
\end{equation}
is satisfied exactly for all rays only for a system with two identical
parabolas producing one-to-one imaging. Here $\alpha$ and
$\alpha^{\prime}$ are angles of the rays to the optical axis emanating
from the object point $O$ and converging to the image point $I$,
respectively; see Fig.~\ref{fig0018a}. In a more general case of a
system with two arbitrary parabolas, the sine condition may be
satisfied to a very good accuracy, in particular, in cases of interest
for x-ray echo spectrometer applications.
 
We consider x~rays reflected from the first parabolic mirror with the
surface given by $x^2=2pz+p^2$, where $p$ is a parabola parameter.
X~rays reflected at a glancing angle of incidence $\gamma$ make an
angle $\alpha=2\gamma$ with the optical axis, see Fig.~\ref{fig0018a},
whose sine is
\begin{equation}
\sin\alpha \,=\, \frac{2px}{x^2+p^2}.
\label{sin010}
\end{equation}
The same ray being reflected from the second mirror with parabolic
surface given by $x^2=2p^{\prime}z+(p^{\prime})^2$ makes an angle
$\alpha^{\prime}$ with the optical axis. The ratio of the sines is
\begin{equation}
\frac{\sin\alpha}{\sin\alpha^{\prime}} = \frac{p}{p^{\prime}}\,\frac{1+\left({p^{\prime}}/{x}\right)^2}{1+\left({p}/{x}\right)^2}.
\label{sin020}
\end{equation}
If the parabolas are identical, i.e., $p=p^{\prime}$, then
${\sin\alpha}/{\sin\alpha^{\prime}}=1$ and the Abbe sine condition is
perfectly fulfilled for all rays, i.e., the system is ``aplanatic''
\cite{BornWolf}, capable of imaging without coma aberrations.

If the parabolas are not identical, i.e., $p\not =p^{\prime}$, but
$p/x\ll 1$ and $p^{\prime}/x\ll 1$, then Eq.~\eqref{sin020} can be
approximated by
\begin{equation}
\frac{\sin\alpha}{\sin\alpha^{\prime}} \simeq \frac{p}{p^{\prime}}\,\left[1+\frac{(p^{\prime})^2-p^2}{x^2}\right].
\label{sin0200}
\end{equation}
For the mirrors with $\gamma=\alpha/2 \simeq 30$~mrad and the focal
lengths, considered in Sec.~\ref{mitigatingcoma}  and Table~\ref{tab-xes}, the typical ratios
are $(p/x)^2 \lesssim 10^{-3}$ at the mirrors' centers, and the variations
are $(p/x)^2 \lesssim 10^{-4}$ over the whole range of $x$ along the
mirrors. 

Therefore, in the particular cases of interest for x-ray echo
spectrometers, the Abbe sine condition for the Wolter-type
parabola-parabola system can be fulfilled with a very good accuracy
even if two different parabolic mirrors are used.



\end{document}